\documentclass[a4paper,11pt]{article}

\usepackage{jheppub} 

\usepackage[T1]{fontenc} 
\usepackage{siunitx}
\usepackage{tikz}
\usepackage{doi}
\usepackage{xcolor}
\definecolor{ds}{RGB}{28,91,247}
\definecolor{tau}{RGB}{28, 140, 240}

\RequirePackage[normalem]{ulem} 
\usepackage{comment}

\title{\boldmath DsTau: Study of tau neutrino production with 400 GeV protons from the CERN-SPS}
\author[a]{Shigeki Aoki,}
\author[b,1]{Akitaka Ariga\note{Corresponding author.},}
\author[b,c]{Tomoko Ariga,}
\author[d]{Sergey Dmitrievsky,}
\author[e]{Elena Firu,}
\author[b]{Dean Forshaw,}
\author[f]{Tsutomu Fukuda,}
\author[d]{Yuri Gornushkin,}
\author[g]{Ali Murat Guler,}
\author[e]{Maria Haiduc,}
\author[h]{Koichi Kodama,}
\author[f]{Masahiro Komatsu,}
\author[g]{Muhtesem Akif Korkmaz,}
\author[i]{Umut Kose,}
\author[e]{Madalina Miloi,}
\author[b]{Antonio Miucci,}
\author[f]{Motoaki Miyanishi,}
\author[f]{Mitsuhiro Nakamura,}
\author[f]{Toshiyuki Nakano,}
\author[e]{Alina Neagu,}
\author[f]{Hiroki Rokujo,}
\author[f]{Osamu Sato,}
\author[d]{Elizaveta Sitnikova,}
\author[f]{Yosuke Suzuki,}
\author[f]{Tomoki Takao,}
\author[d]{Svetlana Vasina,}
\author[b]{Mykhailo Vladymyrov,}
\author[b]{Thomas Weston,}
\author[j]{Junya Yoshida,}
\author[k]{Masahiro Yoshimoto.}

\affiliation[]{The DsTau Collaboration}
\affiliation[a]{Kobe University, Kobe, Japan}
\affiliation[b]{Albert Einstein Center for Fundamental Physics, Laboratory for High Energy Physics, University of Bern, Bern, Switzerland}
\affiliation[c]{Kyushu University, Fukuoka, Japan}
\affiliation[d]{Joint Institute for Nuclear Research, Dubna, Russia}
\affiliation[e]{Institute of Space Science, Bucharest, Romania}
\affiliation[f]{Nagoya University, Nagoya, Japan}
\affiliation[g]{Middle East Technical University, Ankara, Turkey}
\affiliation[h]{Aichi University of Education, Kariya, Japan}
\affiliation[i]{CERN, Geneva, Switzerland}
\affiliation[j]{Advanced Science Research Center, Japan Atomic Energy Agency, Tokai,
Japan}
\affiliation[k]{Gifu University, Gifu, Japan}

\emailAdd{akitaka.ariga@lhep.unibe.ch}

\abstract{
In the DsTau experiment at the CERN SPS, an independent and direct way to measure tau neutrino production following high energy proton interactions was proposed. As the main source of tau neutrinos is a decay of $D_s$ mesons, produced in proton-nucleus interactions, the project aims at measuring a differential cross section of this reaction. The experimental method is based on a use of high resolution emulsion detectors for effective registration of events with short lived particle decays. Here we present the motivation of the study, details of the experimental technique, and the first results of the analysis of the data collected during test runs, which prove feasibility of the full scale study of the process in future.}

\begin{document}

\maketitle
\flushbottom


\section{Introduction}
\label{sec:intro}

Tau neutrino existence was predicted after the tau lepton discovery in 1975 \cite{perl}, and then it was further strengthened by the confirmation of three active neutrinos by LEP in 1989 \cite{aleph,delphi,l3,opal}.
However, only in 2000, 25 years later from the discover of tau lepton, first tau neutrinos were detected in the DONuT experiment \cite{donut0}. In 2008, DONuT published the tau neutrino cross section result with a rather large systematic uncertainty \cite{donut}. In 2015, flavour changing neutrino oscillations were established by detecting appearance signals of neutrino oscillations, including $\nu_\tau$ appearance through $\nu_{\mu,e} \leftrightharpoons \nu_\tau$ oscillations by OPERA \cite{opera} and Super-Kamiokande (SK) \cite{sk_nutau}. 
IceCube \cite{icecube_nutau} also reported an evidence of $\nu_\tau$ appearance recently.

\begin{figure}[htbp]
\centering
\includegraphics[width=0.8\textwidth]{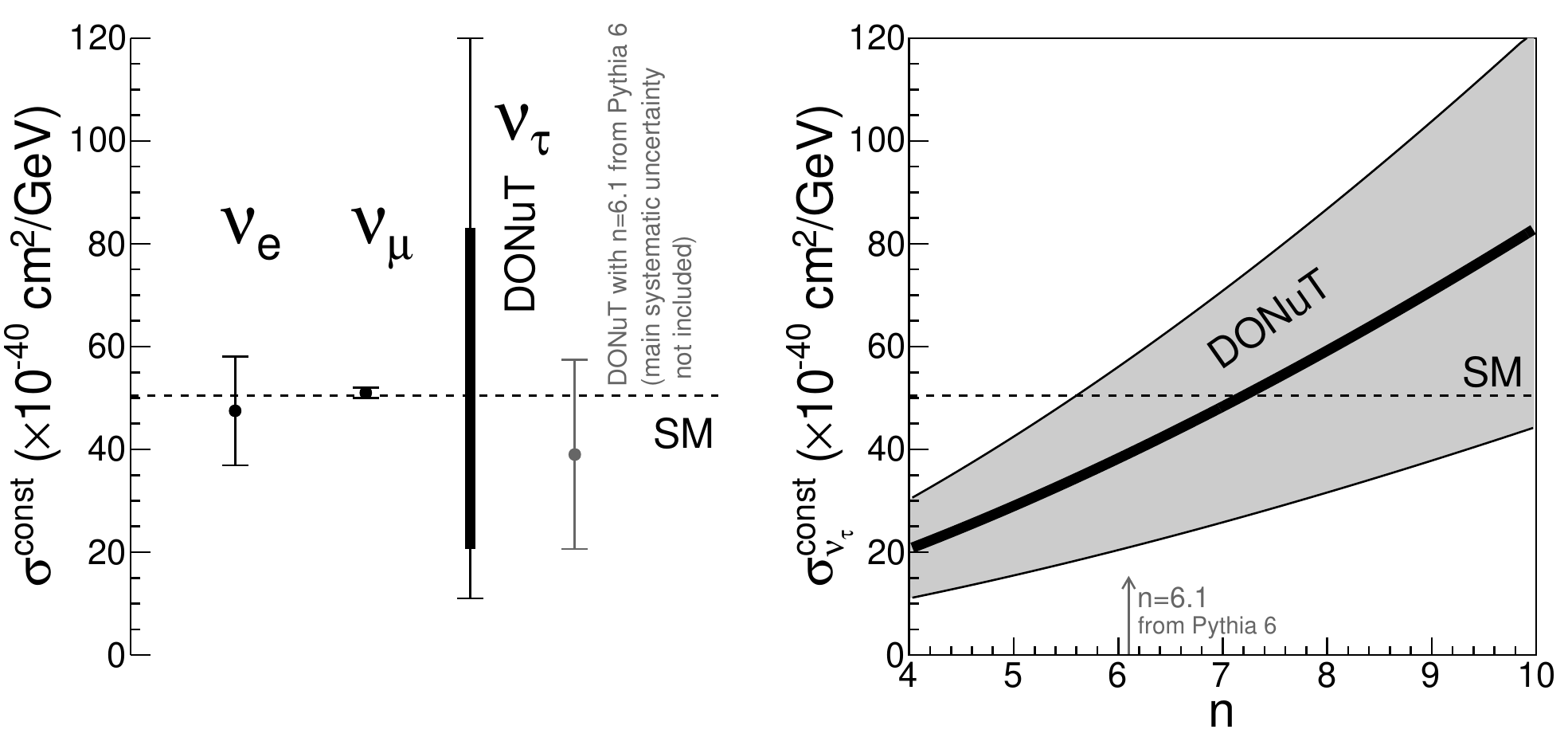}
\caption{Left: $\nu$, $\overline{\nu}$ averaged energy independent cross section of the three neutrino flavors (\cite{gargamelle} for $\nu_e$, \cite{pdg2014} for $\nu_\mu$, and  for $\nu_\tau$ \cite{donut}). The SM prediction is indicated as a dashed horizontal line. For the DONuT result, since there is no measurement of the parameter $n$ concerning the $D_s$ double differential production cross-section (Eq. \ref{eq:differential_crosssection}), the value is plotted in the empirical range of a parameter $n$ given by the DONuT paper as in the right plot. Right: The cross section result by the DONuT experiment, as a function of the parameter $n$ \cite{donut}. The horizontal dashed line corresponds to the predicted value from lepton universality. The arrow shows the value of $n$ estimated from PYTHIA simulation (version 6.129).
}
\label{cross-sec-nu}
\end{figure}

Given the poor measurements of tau neutrinos, their properties  are not well studied. In particular, the cross section of tau neutrino charge current (CC) interaction is known with much larger statistical and systematical uncertainties compared to the other neutrino flavors, as shown in Figure \ref{cross-sec-nu}. On the other hand, a precise measurement of the tau neutrino interaction cross section would be interesting since it could allow testing of the Lepton Universality (LU)  in neutrino scattering.  LU is a principal assumption 
of the Standard Model (SM) of particle physics,  however its validity was questioned by recent results on the B decay asymmetry \cite{babar, lhcb_b+, lhcb_b0}. There is an expectation of the same effect in $\nu_\tau$ scattering \cite{nonuniversality_nutau}. The measurement of the $\nu_\tau$ CC cross section has also a practical impact on current and future neutrino oscillation experiments. Mass hierarchy measurements in the atmospheric Super-Kamiokande (SK) \cite{SK} and accelerator neutrino experiments (for example, in DUNE \cite{dune} and HyperKamiokande \cite{hyper-k}) rely on $\nu_e$ flux measurements, but the sample of registered events is contaminated by $\nu_\tau$ interactions due to $\tau \rightarrow e$ decays. Unlike other error sources, the uncertainty of the $\nu_\tau$ cross section cannot be constrained by near detector measurements. The systematic uncertainty from the $\nu_\tau$ interaction cross section will be a limiting factor in oscillation analyses in these experiments  \cite{wendell, meloni}.   
A better knowledge of the tau neutrino properties would also be a basic input to the astrophysical $\nu_\tau$ observation by IceCube \cite{icecube}.

So far the tau neutrino interaction cross section was only measured in the DONuT \cite{donut}, OPERA \cite{opera} and SK \cite{skxsec} experiments though under rather different conditions.  
DONuT was the first experiment which has directly registered 9 events of the $\nu_\tau$ CC interactions in the accelerator-based neutrino beam and extracted the value of the cross section under certain assumptions on the tau neutrinos flux. 
SK reported the cross section measurement with oscillated atmospheric neutrinos, based on an estimation of amount of tau neutrino events in the total sample by the Neural Network analysis. The energy ranges of the $\nu_\tau$ registered in both experiments are rather different, which makes the results difficult for comparison as different processes are responsible for the tau neutrino scattering. Nevertheless SK has reported on the consistency of the results with the DONuT measurements under several assumptions. 
OPERA registered 10 oscillated tau neutrinos and published the cross section estimation though with a rather large error \cite{operaxsec}. 
In general, all the measurements performed so far have large statistical and/or systematical errors of 30-50\% due to low statistics and experimental uncertainties.
In a future experiment at CERN, SHiP \cite{ship}, it is proposed to perform a rich neutrino program \cite{shipnu} and is expected to collect thousands of tau neutrino interactions, hence providing a negligible statistical error of the cross section measurement. Therefore the overall accuracy of the cross section will be determined by the systematic error, and, in particular, by the $\nu_\tau$ flux uncertainty, which is to be studied by the DsTau experiment.
\par

The dominant source ($>90$\%)  of $\nu_\tau$ in the accelerator-based neutrino beam is leptonic decays of $D_s^\pm$ mesons produced in proton-nucleus interactions:  
\begin{align*}
    D_s^- \rightarrow &\tau^- \overline{\nu}_\tau\\
                      &\tau^-\rightarrow X \nu_\tau,
\end{align*}
producing $\nu_\tau$ and $\overline{\nu}_\tau$ in every decay. 
The tau neutrino beam produced in this way has rather large beam divergence due to the emission angle of $D_s$ and also the decay momenta. Taking into account the finite acceptance of the detectors, it is necessary to know the differential production cross section of $D_s$ to calculate the tau neutrino flux on the detector.  For example, in the case of DONuT only about 6\% of the produced $\nu_\tau$ intersected the DONuT detector, and 5\% are expected in SHiP \cite{ship}.
Conventionally, the differential production cross section of charmed particles is approximated by a phenomenological formula 
 \begin{equation}
     \frac{d^2 \sigma}{dx_F \cdot dp^2_T } \propto (1 - \mid x_F\mid)^n\cdot e^{-b\cdot p^2_T},
\label{eq:differential_crosssection}
 \end{equation} 
 where $x_F$ is the Feynman x ($x_F$ = 2$p^\textnormal{CM}_Z /\surd{s}$ ) and \  $p_T$ is the transverse momentum. 
 $n$ and $b$ are the parameters controlling the longitudinal and transverse dependence of the differential production cross section, respectively.
Although there were several measurements on charm particles \cite{Hera-B, E653, E743, E769, SELEX}, there is a lack of measurements on the $D_s$ differential production cross section in the proton interactions, especially concerning the longitudinal dependence represented by the parameter $n$ (see a detailed review in Appendix \ref{appendix:charm}). This has been the main uncertainty of the $\nu_\tau$ cross section measurements.


At the deep inelastic interaction regime like in DONuT, the tau neutrino cross section, can be approximated as 
\begin{equation}
     \sigma_{\nu_\tau} = \sigma^\textnormal{const}_{\nu_\tau} \cdot E \cdot K(E)
\end{equation}
where  $\sigma^\textnormal{const}_{\nu_\tau}$ is an energy independent cross section, $E$ - the neutrino  energy and $K(E)$ - a factor taking into account kinematic effects due to the $\tau$-lepton mass.
The lack of experimental constraints on the parameter $n$ relating to the $D_s$ differential production cross section forced the DONuT collaboration to present their result as a function of $n$, as
\begin{equation}
    \sigma^\textnormal{const}_{\nu_\tau} = 2.51 n^{1.52}  (1\pm0.33(\textnormal{stat}) \pm 0.33 (\textnormal{syst}))\times 10^{-40} \  \textnormal{cm}^2\ \textnormal{GeV}^{-1}.
\end{equation}
Figure \ref{cross-sec-nu} on the right shows this DONuT result plotted in the empirical range of $n$.

DONuT also presented the numerical value for the cross section by making use of the value $n=6.1$ derived from the PYTHIA simulation \cite{pythia61} (indicated in Figure \ref{cross-sec-nu}), as 
\begin{equation}
    \sigma^\textnormal{const}_{\nu_\tau}=(0.39\pm 0.13 (\textnormal{stat}) \pm 0.13 (\textnormal{sys})) \times 10^{-38}\ \textnormal{cm}^2\ \textnormal{GeV}^{-1}.
\end{equation}
Note that the contribution due to the uncertainty of $n$ is omitted from this formula. When it is included, the relative systematic uncertainty exceeds 50\%. 

Thus, a new measurement of differential production cross section of $D_s$ is vital for future neutrino experiments as well as the re-evaluation of the DONuT result.

 

	In the DsTau experiment, a direct study of tau neutrino production, namely a measurement of $D_s \rightarrow \tau \rightarrow X$ decays following high-energy proton-nucleus interactions, 
	is proposed. 
	DsTau exploits a simple setup consisting of a segmented high resolution vertex detector capable to recognize $D_s \rightarrow \tau\rightarrow X$ by their very peculiar double-kink topology. The project aims to detect $\sim 1000$ \  $D_s \rightarrow \tau\rightarrow X$  decays in 2.3$\times$10$^8$ proton interactions with the tungsten target. State-of-the-art nuclear emulsion detectors with a nanometric-precision readout will be used to achieve this goal.
	The modern use of the emulsion detection technology is based on the fast evolution of the high-speed and high-precision automatic readout of emulsions developed during the last two decades and available today \cite{hts1, hts2, atmic}.
	
	DsTau will provide an independent  $\nu_\tau$ flux prediction for future neutrino beams with accuracy under 10\%. Then, the systematic uncertainty of the $\nu_\tau$ CC cross section measurement can be made sufficiently low to test LU in neutrino scattering.

In addition to the primary aim of measuring $D_s$ differential production cross section, in 2.3$\times10^8$ proton interactions, a high yield of $\mathcal{O}(10^5)$ charmed particle pairs is expected. The analysis of those events can provide valuable by-products, 
such as a measurement of the intrinsic charm content in proton \cite{intrinsic_charm} by measuring the emission angle (pseudorapidity) of the charmed particle pairs, the interaction length of charmed hadrons, the $\Lambda_c$ production rate and search of super-nuclei \cite{supernuclei}, that have never measured.

\section{Overview of project}
\label{overview}

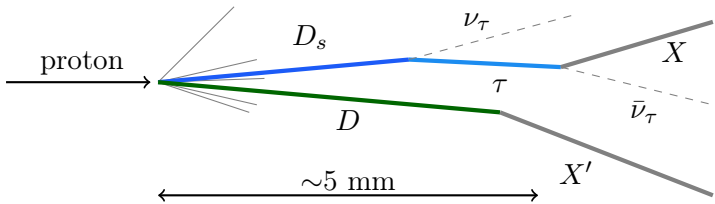
\begin{figure}[htbp]
\centering
\begin{tikzpicture}[xscale=1.0, every node/.style={transform shape}]
\coordinate (V) at (0,0);
\coordinate (Ds) at (3.3,0.3);
\coordinate (tau) at (5.3,0.2);
\coordinate (nutau1) at (5.5,0.9);
\coordinate (nutau2) at (7.3,-0.3);
\coordinate (X) at (7.3,0.8);
\coordinate (D) at (4.5,-0.4);
\coordinate (XX) at (7.3,-1.5);
\coordinate (proton) at (-2,0);

\draw[gray] (V) -- (1,1);
\draw[gray] (V) -- (1.4,0.05);
\draw[gray] (V) -- (1.3,0.3);
\draw[gray] (V) -- (1.3,-0.3);
\draw[gray] (V) -- (1.2,-0.4);
\draw[->,thick,black] (proton) -- (-0.1,0);
\draw[ultra thick, ds] (V) -- (Ds);
\draw[ultra thick, tau] (Ds) -- (tau);
\draw[gray, dashed] (Ds) -- (nutau1);
\draw[gray, dashed] (tau) -- (nutau2);
\draw[ultra thick, black!60!green] (V) -- (D);
\draw[ultra thick, gray] (D) -- (XX);
\draw[ultra thick, gray] (tau) -- (X);
\draw[<->,thick, black] (0,-1.5) -- (5, -1.5);

\node [] at (-1, 0.25) {proton};
\node [] at (2, 0.6) {$D_s$};
\node [] at (4.2, 0.8) {$\nu_\tau$};
\node [] at (4.5, -0.05) {$\tau$};
\node [] at (6.8, 0.4) {$X$};
\node [] at (6.4, -0.4) {$\bar{\nu}_\tau$};
\node [] at (5.5, -1.2) {$X'$};
\node [] at (2.5, -0.5) {$D$};
\node [] at (2.5, -1.3) {$\sim$5 mm};

\end{tikzpicture}

\caption{Topology of $D_s\rightarrow\tau\rightarrow X$ events.}
\label{topology}
\end{figure}

The topology of  $D_s \rightarrow \tau \rightarrow X$  events appears as a ``double-kink'', as shown in Figure \ref{topology}. 
In addition, because charm quarks are created in pairs, another decay of a charged/neutral charmed particle will be observed with a flight length of a few millimetres.
 Such a double-kink plus decay topology in a short distance is a very peculiar signature of this process, and the background mimicking this topology is marginal. 

\begin{figure}[htbp]
\begin{center}
\includegraphics[height=5cm]{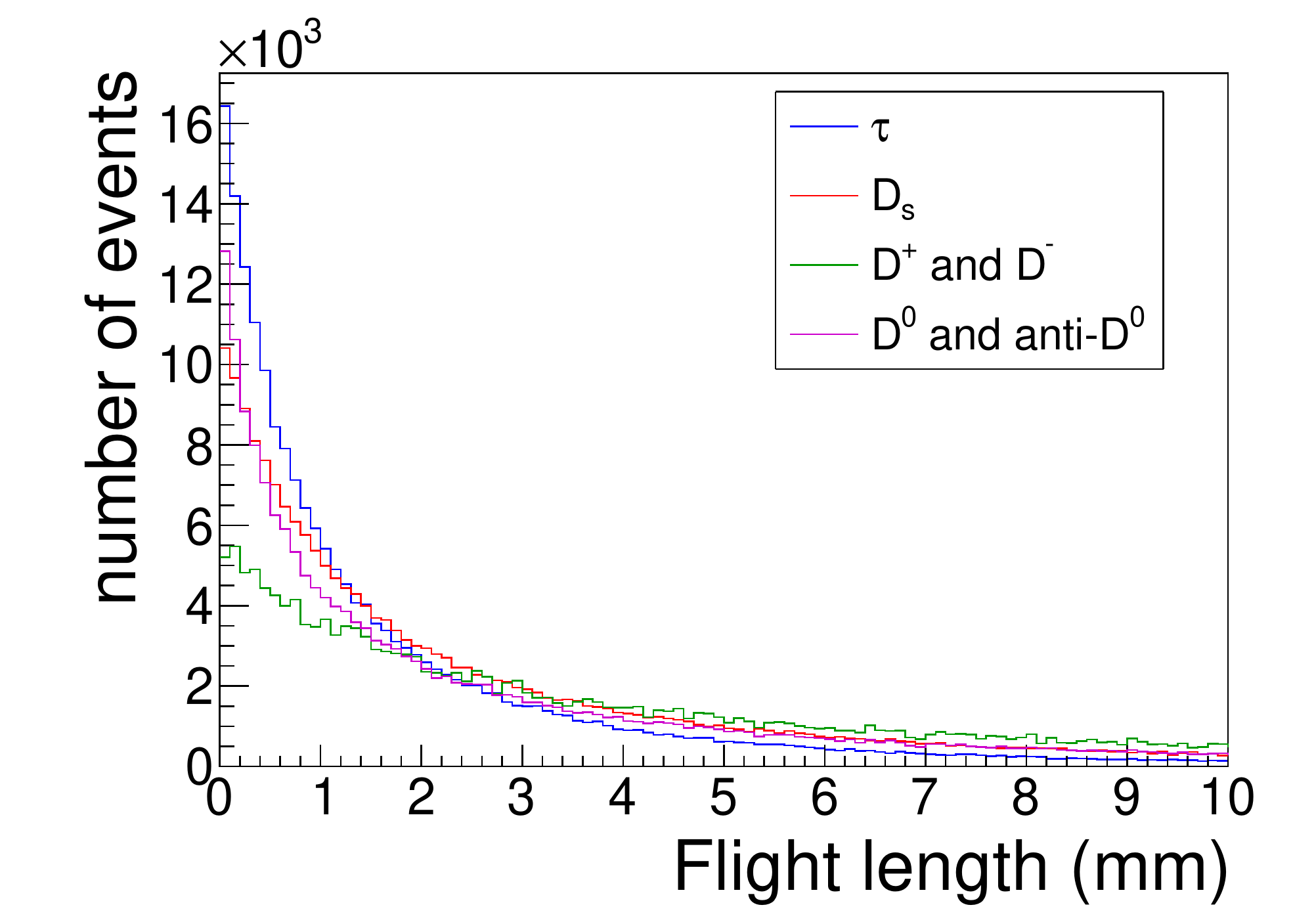}
\includegraphics[height=5cm]{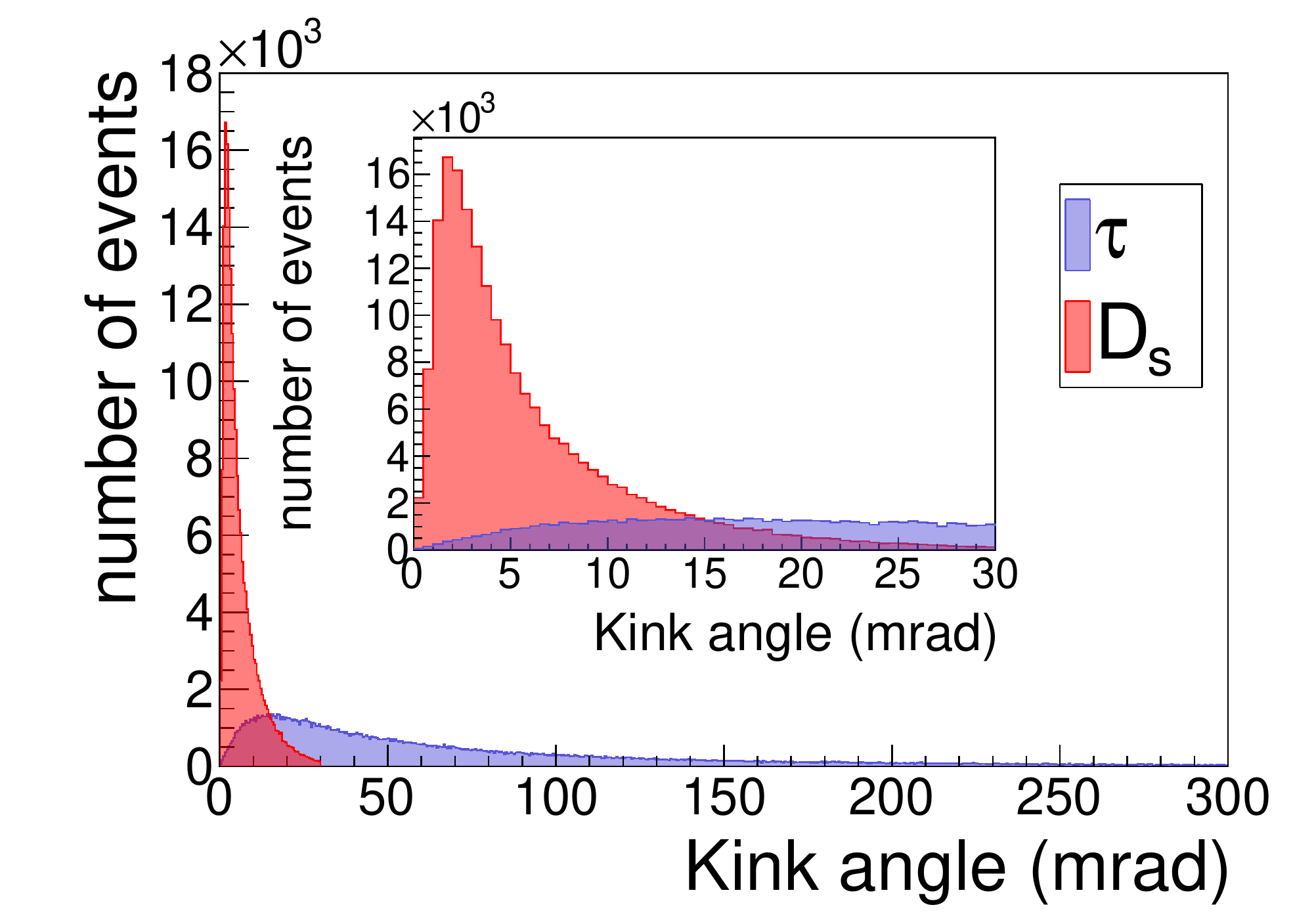}
\caption{Expected distributions of flight length (left) and kink angle (right) of charmed particles and $\tau$. }
\label{fig:fl_kink}
\end{center}
\end{figure}

However, to register the events is a challenge.  First, all the decays take place at a scale of millimetres. Second, the kink angle of $D_s \rightarrow \tau$ is anticipated to be very small. The expected signal features were studied making use of Pythia 8.1 \cite{pythia81}, as shown in Figure \ref{fig:fl_kink}. The mean flight lengths of $D_s$, $\tau$ and pair-charms are 3.6 mm, 2.1 mm and 4.2 mm, respectively. Although the kink angle at $\tau$ decays vertex is easily recognizable (mean kink angle of 96 mrad), the kink angle at $D_s\rightarrow\tau$ decays is rather small, 6.2 mrad on average.  The measurement of $D_s$ momentum is also difficult as two 
$\nu_\tau$ escape measurement, spoiling the invariant mass reconstruction. Therefore we'll use an emulsion detector with nanometric spatial accuracy to efficiently recognize the events.

\begin{figure}[htbp]
\begin{center}
 \includegraphics[width=\textwidth]{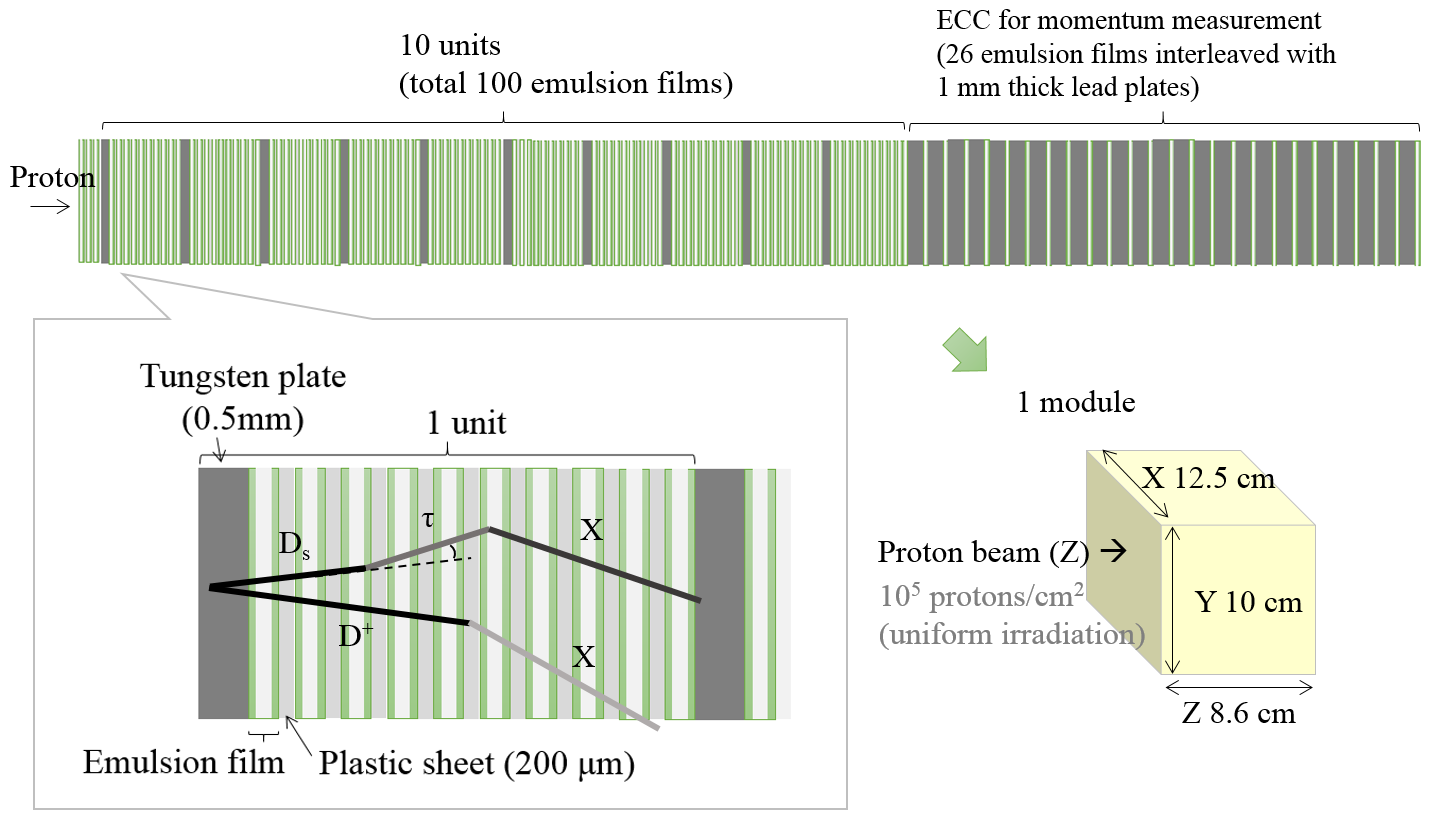}
\caption{Schematic view of the module structure. A tungsten target plate, is followed by 10 emulsion films alternated by 9 plastic sheets acting as a tracker and decay volume of 4.8 mm. The sensitive layers of emulsion detectors are indicated by green color. This basic structure is repeated 10 times, and then followed by a lead-emulsion ECC structure for momentum measurement of the daughter particles.}
\label{module}
\end{center}
\end{figure}

The DsTau module structure is shown in Figure \ref{module}. The structure is divided into two parts. The upstream part is named the \textit{decay module}. The basic unit is made of a 500-\si{\micro m}-thick tungsten plate (target) followed by 10 emulsion films interleaved with 9 200-\si{\micro m}-thick plastic sheets which act as a decay volume for short-lived particles as well as high-precision particle trackers. This structure (thickness of 5.3 mm) is repeated 10 times. Five additional emulsion films are placed most upstream of the module to tag the incoming beam protons. 
It is then followed by the downstream part
made of a repeated structure of emulsion films and 1-mm-thick lead plates for momentum measurement of the daughter particles.
The entire detector module is 12.5 cm wide, 10 cm high and 8.6 cm thick and consists of a total of 131 emulsion films.
Momentum of the particles can be determined by Multiple Coulomb Scattering (MCS) measurement along their tracks in the tungsten plates and in the downstream ECC \cite{mcs}.

\begin{figure}[htbp]
\begin{center}
\includegraphics[width=\textwidth]{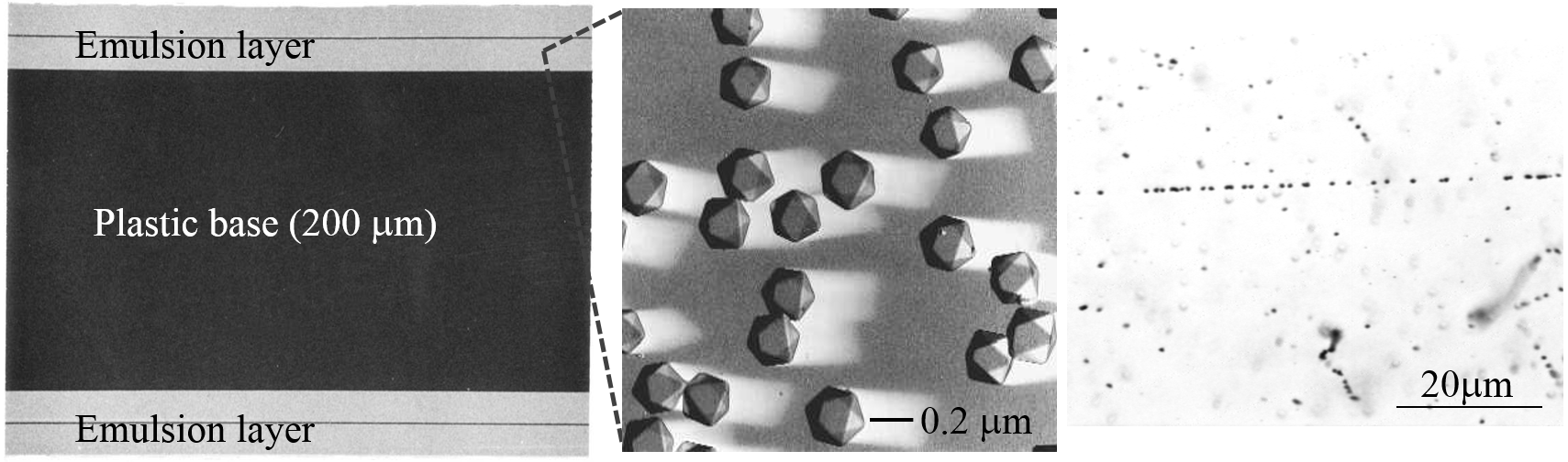}
\caption{A cross-sectional view of an emulsion film \cite{opera_film} (left), electron microscope picture of silver halide crystals (middle) and a
track of a relativistic particle (10 GeV/c $\pi$ beam) (right).}
\label{emulsion_film}
\end{center}
\end{figure}

The emulsion detector (Figure \ref{emulsion_film}) comprises a countless number of silver halide crystals dispersed in the gelatin media. Each of those crystals is a semiconductor with a band gap of 2.5 eV and works as an individual detection channel for charged particles. 
The high density of detection channels, of the order of $10^{14}$ channels/\si{cm^3}, makes this detector suitable for three-dimensional high-precision measurements. 

The emulsion films for DsTau have two emulsion layers (70 \si{\micro m} thick) poured onto both sides of a 210-\si{\micro m}-thick plastic base
\footnote{In the 2016/2017 test runs, 50-\si{\micro m}-thick emulsion layers and 180-\si{\micro m}-thick plastic base were used.}.  The diameter of the silver halide crystals is ~200 nm, as shown in Figure \ref{emulsion_film} (middle). 
Once a charged particle passes through the emulsion layer, the ionization is recorded quasi-permanently, and then amplified and fixed by the chemical process. A trajectory of a 
charged particle can be observed under an optical microscope as shown in Figure \ref{emulsion_film} (right).
An emulsion detector with 200-nm-diameter crystals and a 210-\si{\micro m}-thick base has a track position resolution of 50 nm \cite{intrinsic_resolution} and an angular resolution of 0.34 mrad (projection). With this angular resolution, one can detect 2-mrad kink with 4$\sigma$ confidence.

A key feature of the modern emulsion technique is the use of the fast readout instruments, which allow extracting and digitizing the information on the tracks fully automatically. Emulsion detectors and the automated readout systems have been successfully employed by several neutrino experiments such as CHORUS \cite{chorus}, DONuT  \cite{donut0, donut} and OPERA \cite{opera}.
The latest scanning system, the HTS system \cite{hts1,hts2}, allows scanning of the emulsion films at the speed of 5,000 \si{cm^2} per hour per emulsion layer, which is $\mathcal{O}(100)$ faster than those used in OPERA. The high performance is achieved by combining a custom-made wide-field objective lens and 72 high-speed image sensors. The image data is processed in real-time manner by a dedicated track recognition software based on the GPU technology. An effective scanning performance reached a 1000 \si{m^2}/year.

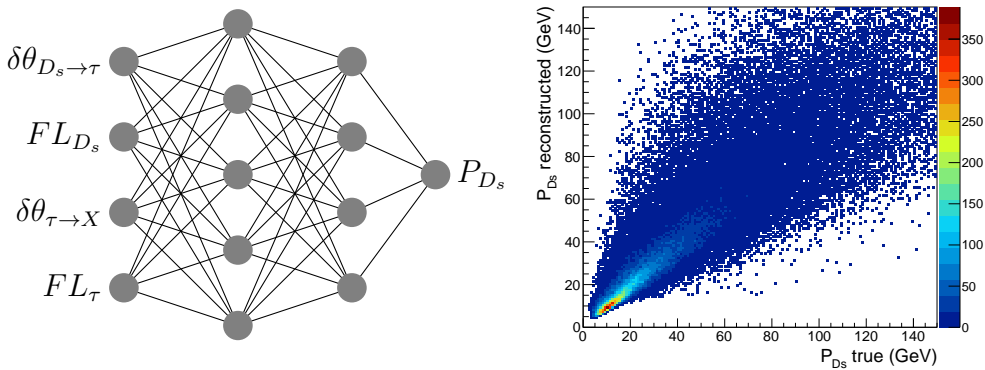
\begin{figure}[htpb]
\centering
\begin{tabular}{cc}

\begin{tikzpicture}[every node/.style={transform shape}]
\large
\def\x{0};
\def\xx{1.5};
\def\xxx{3};
\def\xxxx{4.1};

\node [anchor=east ] (p11) at (0, 5) {$\delta\theta_{D_s\rightarrow \tau}$ \ };
\node [anchor=east ] (p12) at (0, 4) {$FL_{D_s}$ \ };
\node [anchor=east ] (p13) at (0, 3) {$\delta\theta_{\tau\rightarrow X}$ \ };
\node [anchor=east ] (p14) at (0, 2) {$FL_\tau$ \ };
\node [anchor=west ] (p41) at (\xxxx, 3.5) {\ $P_{D_s}$};

\node [anchor=west ] at (\xxxx, 1) {};
\def\ymin{3.5};

\typeout{hoge}
\foreach \y in {5,4,3,2}
\foreach \yy in {5.5,4.5,3.5,2.5,1.5}
\draw[thin, black] (\x,\y) -- (\xx,\yy);

\foreach \yy in {5.5,4.5,3.5,2.5,1.5}
\foreach \y in {5,4,3,2}
\draw[thin, black] (\xx,\yy) -- (\xxx,\y);

\foreach \y in {5,4,3,2}
\draw[thin, black] (\xxx,\y) -- (\xxxx,3.5);

\foreach \y in {5,4,3,2} \node[circle,color=black,fill=gray] at (\x,\y) {};
\foreach \y in {5.5,4.5,3.5,2.5,1.5} \node[circle,color=black,fill=gray] at (\xx,\y) {};
\foreach \y in {5,4,3,2} \node[circle,color=black,fill=gray] at (\xxx,\y) {};
\node[circle,color=black,fill=gray] at (\xxxx,3.5) {};
\end{tikzpicture}
\includegraphics[width=0.4\textwidth]{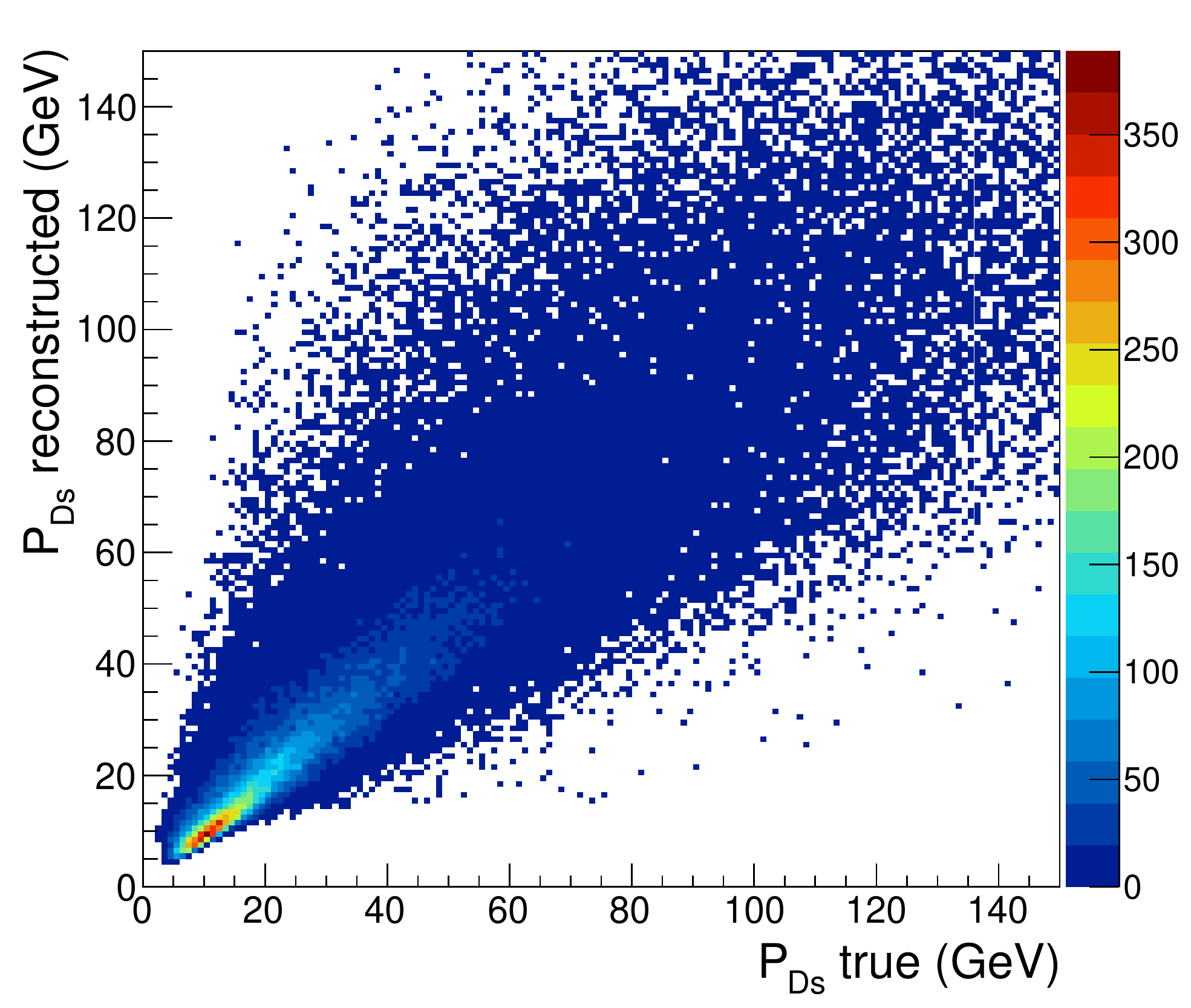}
\end{tabular}
\caption{Left: Design of a neutral network for reconstruction of the $D_s$ momentum using the topological variables $\theta_{D_s\rightarrow\tau}$, $\theta_{\tau\rightarrow X}$, $FL_{D_s\rightarrow\tau}$, $FL_{\tau\rightarrow X}$. Right: Reconstructed $D_s$ momentum versus its true value. A Gaussian fit of $\Delta P/P$ provides $\sigma$ of 20\%.
}
\label{fig:dsmom}
\end{figure}

The $D_s$ momentum ($P_{D_s}$) will be reconstructed by means of a machine-learning algorithm using topological variables of the event. There are two flight lengths ($FL_{D_S}$, $FL_{\tau}$) and two kink angles ($\delta\theta_{D_s\rightarrow\tau}$, $\delta\theta_{\tau\rightarrow X}$), those are proportional and inverse proportional to the $\gamma$ factor of decaying particles. The combination of these four variables effectively provides an estimate of $P_{D_s}$. The algorithm was trained and tested with the simulated sample ($D_s \rightarrow \tau \rightarrow$ 1 charged particle)
(Figure~\ref{fig:dsmom}). The momentum resolution is estimated to be 20\%.


\begin{figure}[htbp]
\begin{center}
\begin{tabular}{cc}
\includegraphics[height=5cm]{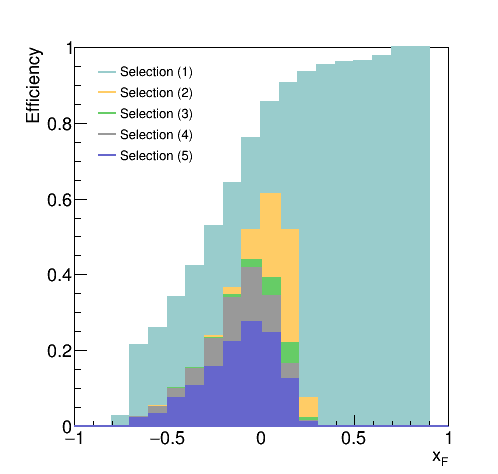}&
\includegraphics[height=5cm]{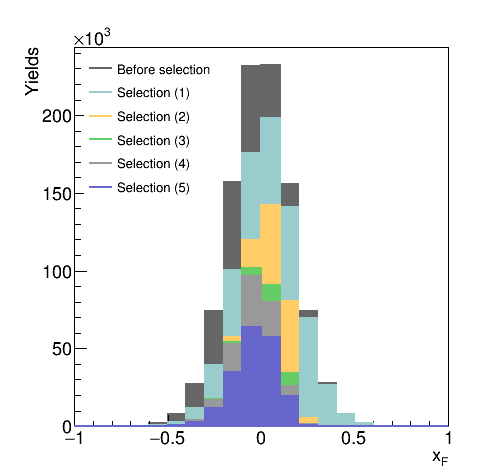}
\end{tabular}
\caption{The estimated detection efficiency as a function of $x_F$ (left) and the $x_F$ distributions after the selection (right). Selection (1)-(5) are described in the text.
}
\label{fig:eff}
\end{center}
\end{figure}

\begin{figure}[htbp]
\begin{center}
\begin{tabular}{cc}
\includegraphics[height=5cm]{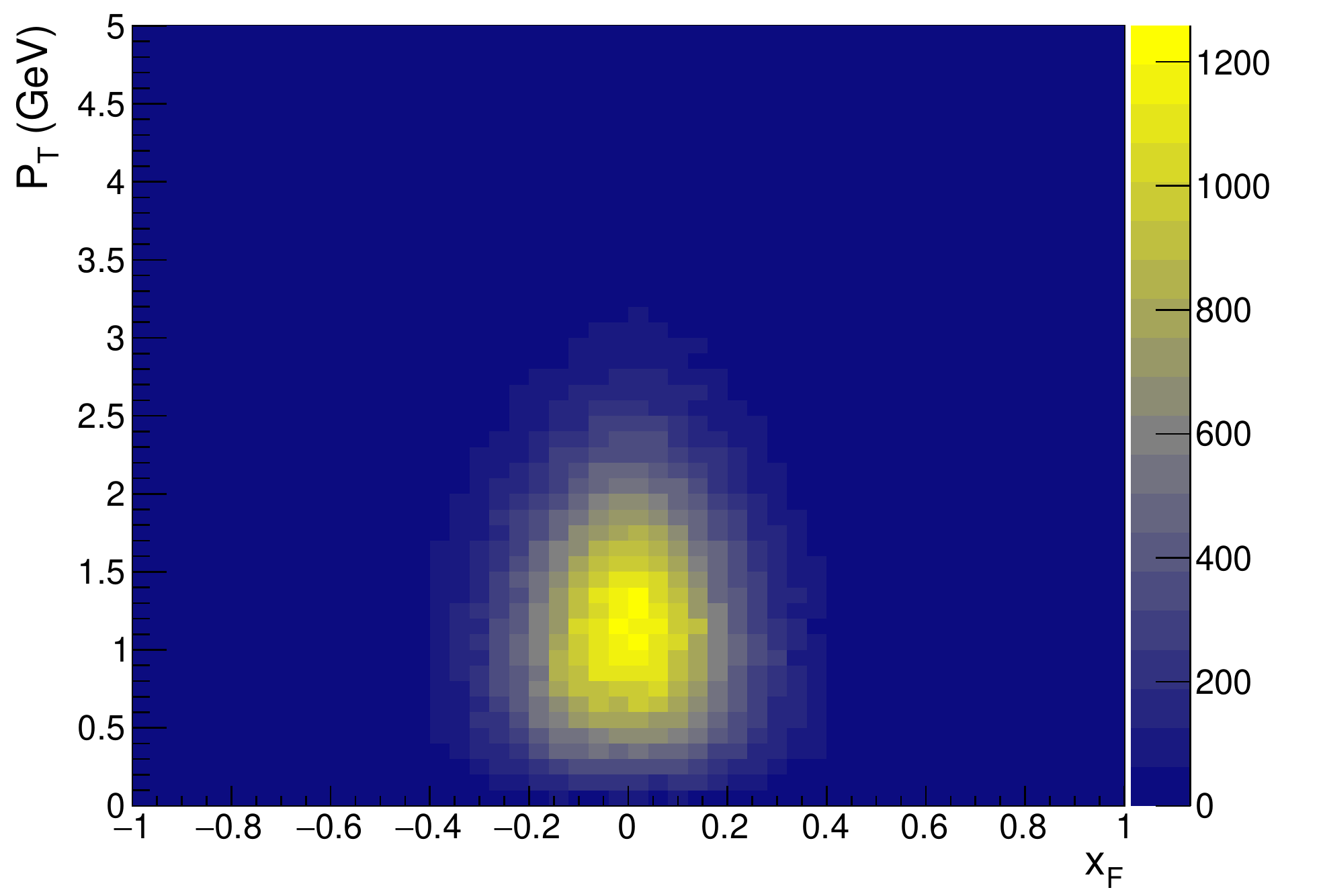}&
\includegraphics[height=5cm]{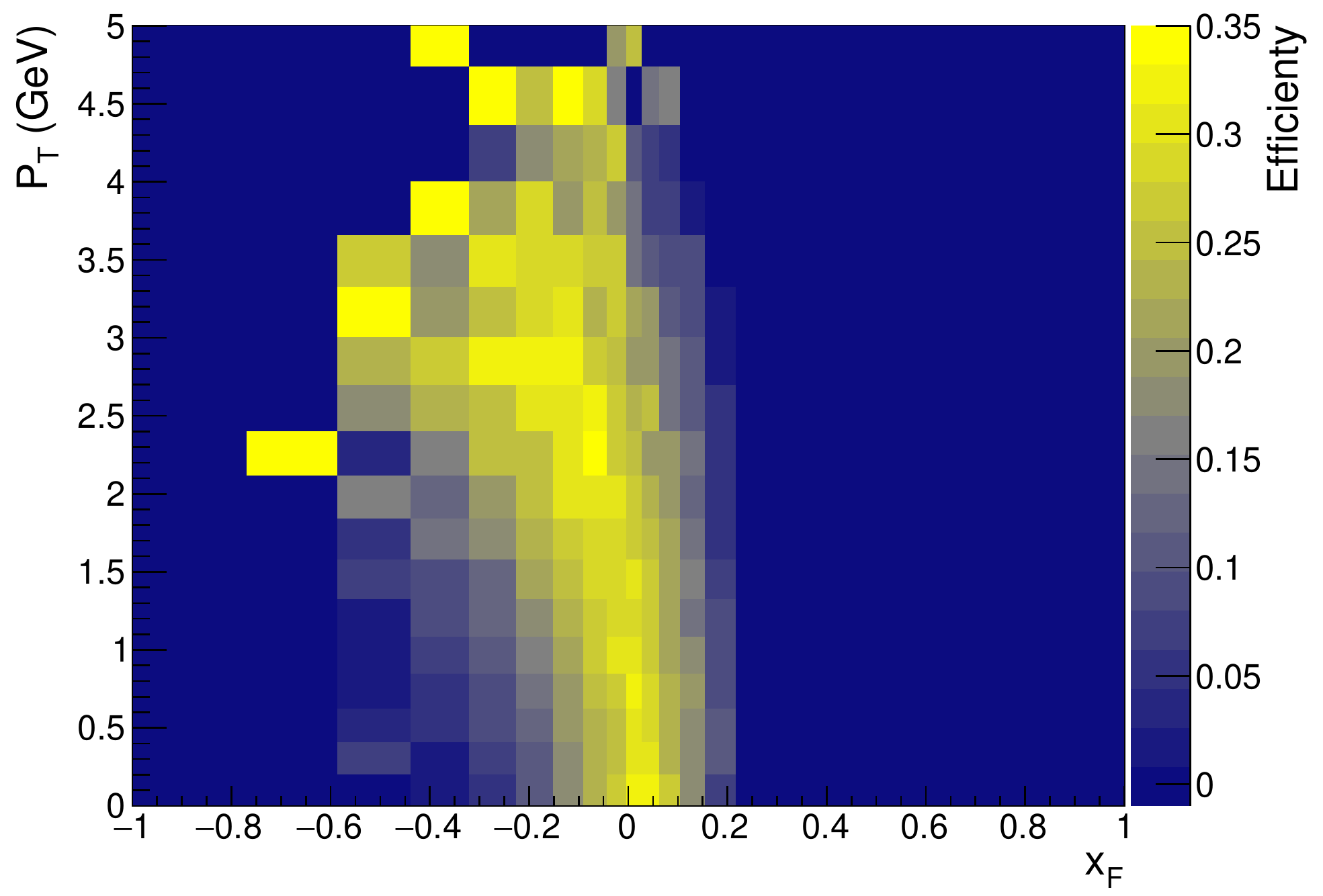}
\end{tabular}
\caption{The $x_F$-$p_T$ distribution before the selection (left) and the estimated detection efficiency in the $x_F$-$p_T$ plane (right).
}
\label{fig:xf-pt}
\end{center}
\end{figure}

The detection efficiency for the $D_s \rightarrow \tau \rightarrow 1$ prong events (85\% of $\tau$ decays) is estimated by PYTHIA 8.1 \cite{pythia81} simulation. 
Table \ref{tb:eff} shows the selection criteria and breakdown of the estimated efficiency. The overall registration efficiency of the reaction is estimated to be 20\%.
Figure \ref{fig:eff} shows the estimated detection efficiency as a function of the Feynman $x$ ($x_F=2p^{CM}_Z/\sqrt{s}$) (left) and the $x_F$ distributions after the selection (right). Figure \ref{fig:xf-pt} shows the $x_F$ - $p_T$ distribution before the selection (left) and the detection efficiency in the $x_F$ - $p_T$ plane (right). The efficiency for high $x_F$ component falls by Selection (2) concerning the kink angle $\Delta\theta_{D_s\rightarrow\tau}\geq$ 2 mrad, for which a further improvement of the detection efficiency is under study. 
Since the angular resolution depends on the track length, it is expected to be better 
for the particles passing through more than two sensitive layers (see also Figure \ref{fig:alignment}). So, the efficiency can be further improved by applying the varying threshold for the first kink angle, depending on the flight length of the parent and daughter.

\begin{table}[hbtb]
\begin{center}
\begin{tabular}{lr}
\hline
Selection & Efficiency (\%)\\
\hline
(1) Flight length of $D_s$ $\geq$ 2 emulsion layers &  77\\
\hline
(2) Flight length of $\tau$ $\geq$ 2 emulsion layers and $\Delta\theta_{D_s\rightarrow\tau}\geq$ 2 mrad & 43 \\
\hline
(3) Flight length of $D_s$ $<$ 5 mm and flight length of $\tau$ $<$ 5 mm & 31 \\
\hline
(4) $\Delta\theta_{\tau \rightarrow X}\geq$ 15 mrad & 28 \\
\hline
(5) Pair charm: 0.1 mm $\leq$ flight length $<$ 5 mm & 20 \\
(charged decays with $\Delta\theta\geq$ 15 mrad or neutral decays) & \\
\hline
\end{tabular}
\caption{
Breakdown of the efficiency estimation for $D_s\rightarrow \tau \rightarrow 1$-prong decay.}
\label{tb:eff}
\end{center}
\end{table}

\par The main background to $D_s \rightarrow \tau \rightarrow 1$ prong events are 
hadron interactions that can mimic decays of short lived particles.
The probability of background event appearance is calculated as follows,
\[
P_\textnormal{BG}^{D_s\rightarrow \tau \rightarrow\textnormal{1 prong + pair charm}}        =    P_\textnormal{int} \cdot 
    P_\textnormal{BG}^{D_s\rightarrow\tau}\cdot
    P_\textnormal{BG}^{\tau \rightarrow\textnormal{1 prong}}\cdot
    P_\textnormal{BG}^{\textnormal{pair charm}}.
\]
Here, $P_\textnormal{int}$ is the probability of a primary proton to interact in the tungsten target in the decay module. $P_\textnormal{BG}^{D_s\rightarrow\tau}$, $P_\textnormal{BG}^{\tau \rightarrow\textnormal{1 prong}}$ and $P_\textnormal{BG}^{\textnormal{pair charm}}$ are the probabilities of secondary hadron interactions to be identified as $D_s \rightarrow \tau$, $\tau \rightarrow \textnormal{1 prong}$ and pair charm decays, respectively.

\par These probabilities are obtained by simulating $3\times 10^5$ protons on the detector with the FLUKA~\cite{Fluka1, Fluka2} simulation tools. The criteria, used for charged charm or tau decay topology selection, are applied to interactions of secondary hadrons with only one charged daughter particle ($P>2$ GeV/c). In addition to high energy particles a large part of interactions has associated nuclear fragments, which is strong evidence of hadron interactions. Those are effectively rejected by requesting only one charged daughter.


\par The sources of background to neutral pair charm appearance are neutral hadron interactions and charged particle interactions within 2 emulsion plates from the proton interaction vertex in case such short tracks are not reconstructed. These backgrounds can be associated with any proton interactions occurred upstream.
To estimate the background probability to neutral charm topology, neutral particle interactions and charged particle interactions within 2 emulsion plates were selected and tested with every primary proton interaction vertex occurred upstream in tungsten.
The selection criteria in Table \ref{tb:eff} were applied to each primary-secondary interactions couple.
Search for nuclear fragments associated with secondary interaction allows suppressing background in this channel as well. 


\par The probability for an incident proton to interact in tungsten and the probabilities of background event appearance with a certain path length and kink angles $\theta_\textnormal{kink}$ and opening angle between daughters $\theta_\textnormal{open}$ used for different signal channels are shown in Table \ref{tab:bg}. The total probabilities to register background events to double kink with charged pair charm or with neutral pair charm are $1.3\pm 0.4\times 10^{-9}$ and $2.7\pm 0.8\times 10^{-9}$ per incident proton, respectively. The expected numbers of background events in the full statistics of DsTau ($4.6\times 10^9$ p.o.t.) are  $6.0\pm1.8$ and $12.4 \pm 3.7$ for these 2 signal channels, respectively.

\begin{table}[!htb]
    \centering
    \small
\begin{tabular}{|l|c|l|}
\hline
Category & Criteria & Probability \\
\hline
    $P_\textnormal{interaction}$&
    interactions in the decay module &
    $(5.00 \pm 0.04) \times 10^{-2}$\\
\hline
    $P_\textnormal{BG}^{D_s\rightarrow\tau}$&
    \begin{tabular}{c}
    2 mrad $<$ $\theta_\textnormal{kink} <$ 30 mrad\\
    0.1 mm $<$ path length $<$ 5 mm 
    \end{tabular} & 
    {$(0.97\pm 0.25)\times 10^{-3}$}\\
\hline
    $P_\textnormal{BG}^{\tau \rightarrow\textnormal{1 prong}}$, $P_\textnormal{BG}^{\textnormal{charged charm}}$ & 
    \begin{tabular}{c}
    20 mrad $<$ $\theta_\textnormal{kink} <$ 500 mrad\\
    0.1 mm $<$ path length $<$ 5 mm 
    \end{tabular} & 
    {$(5.1\pm0.6)\times 10^{-3}$}\\ 
\hline
    $P_\textnormal{BG}^\textnormal{neutral charm}$ & 
    
    \begin{tabular}{c}
    $\theta_\textnormal
    {open} >$ 10 mrad \\
    0.1 mm $<$ path length $<$ 5 mm
    \end{tabular} 
    \color{black}
        & 
    {$(10.9\pm 0.9)\times 10^{-3}$}\\ 
\hline
\hline
    $P_\textnormal{BG}^{D_s\rightarrow \tau \rightarrow\textnormal{1 prong + charged charm}}$ & 
    $P_\textnormal{int.}\cdot P_\textnormal{BG}^{D_s\rightarrow\tau}\cdot
    P_\textnormal{BG}^{\tau \rightarrow\textnormal{1 prong}}\cdot
    P_\textnormal{BG}^{\textnormal{charged charm}}$
    & 
   {$(1.3 \pm 0.4)\times 10^{-9}$}\\ 
\hline
    $P_\textnormal{BG}^{D_s\rightarrow \tau \rightarrow\textnormal{1 prong + neutral charm}}$ & 
    $P_\textnormal{int.}\cdot P_\textnormal{BG}^{D_s\rightarrow\tau}\cdot
    P_\textnormal{BG}^{\tau \rightarrow\textnormal{1 prong}}\cdot
    P_\textnormal{BG}^{\textnormal{neutral charm}}$
    & 
   {$(2.7 \pm 0.8)\times 10^{-9}$}\\ 
\hline
\end{tabular}
    \caption{Probability of hadron interactions mimicking the specific decay topology. $P_\textnormal{BG}^{D_s\rightarrow\tau}$, $P_\textnormal{BG}^{\tau \rightarrow\textnormal{1 prong}}$, $P_\textnormal{BG}^{\textnormal{charged charm}}$, $P_\textnormal{BG}^\textnormal{neutral charm}$ are the probabilities per proton interactions. The rest are those per incident proton. The errors are due to the statistics of the MC data set.}
    \label{tab:bg}
\end{table}

\begin{figure}[htbp]
\begin{center}
\begin{tabular}{cc}
\includegraphics[height=5.2cm]{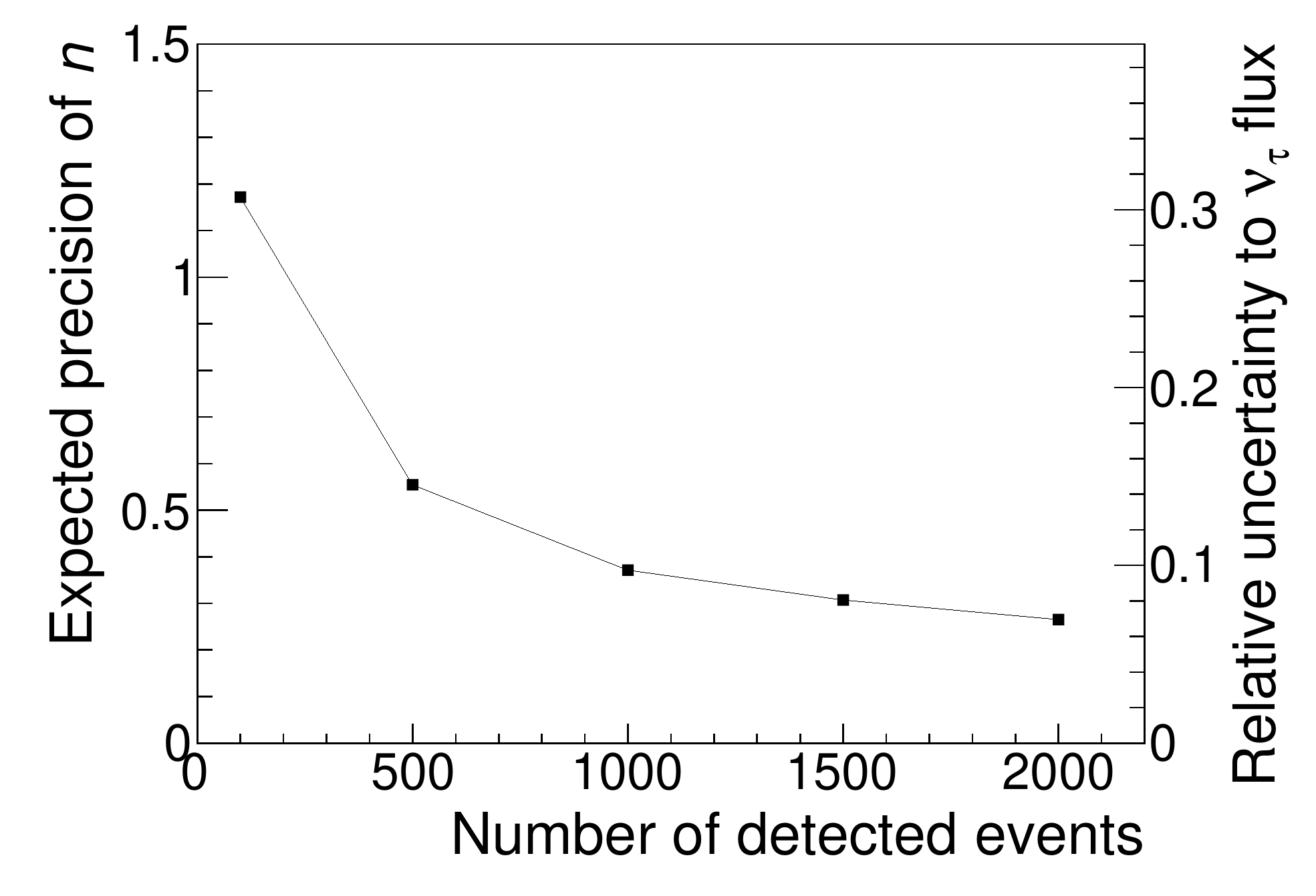}
\end{tabular}
\caption{
Expected precision for the measurement of parameter $n$ as a function of the number of detected $D_s\rightarrow\tau\rightarrow X$ events. The estimated relative uncertainty of the $\nu_\tau$ flux is also given as the y-axis on the right.
}
\label{precision_n}
\end{center}
\end{figure}

DsTau will provide a differential cross section of $D_s$ meson production in 400 GeV proton-nucleus interaction in 2D space, e.g. $x_F-P_\textnormal{T}$ or $P-P_\textnormal{T}$, so that future neutrino experiments can directly implement it.
It may also be fit with the phenomenological formula, Eq. \ref{eq:differential_crosssection}, and the parameter $n$ is estimated, which is relevant for a re-evaluation of the tau neutrino cross section measurement by the DONuT experiment.
The expected precision of the parameter $n$ as a function of available statistics is calculated with a toy MC method and given in Figure 
\ref{precision_n}. At the statistics of 1000 $D_s\rightarrow\tau\rightarrow X$ detected events, the relative uncertainty of the $\nu_\tau$ flux will be reduced to below 10\%.

In order to collect 1000 $D_s \rightarrow \tau$ events, a 230 millions of proton interactions are to be analyzed, which is another challenge from point of view of the track density and amount of data to be processed.
The high proton density of \si{10^5/cm^2} at the upstream surface of emulsion detector was chosen to maximize the number of interactions in a single module. The track density will then increase in the detector, yet not exceeding \si{10^6/cm^2} at the downstream part of the decay module, which is affordable for the emulsion detector readout and reconstruction. With this density, 6.25$\times 10^5$ proton interactions are expected in the tungsten target in a decay module. To accumulate $2.3 \times 10^8$ proton interactions in the tungsten plates,  $4.6 \times 10^9$ protons on target are needed. More than 368 modules with a total film area of 593 \si{m^2} will be employed for this measurement.


\section{Beam exposure and analysis scheme}
\subsection{Data taking}
Two test beam campaigns were held at CERN SPS in 2016 and in 2017. In 2018, a pilot run was conducted aiming at recording 10\% of the experimental data. Here we present first results of the beam tests in 2016 and 2017. 

The emulsion gel was made of 200-nm-diameter silver bromide crystals, whose volume occupancy were 40\% at the dried state. The gel was designed by Nagoya University and a consignment production was done by FUJIFILM in Japan. 
15 \si{m^2} of emulsion films were then produced at the special facility in the University of
Bern. The detector modules were assembled at CERN, right before the exposure.

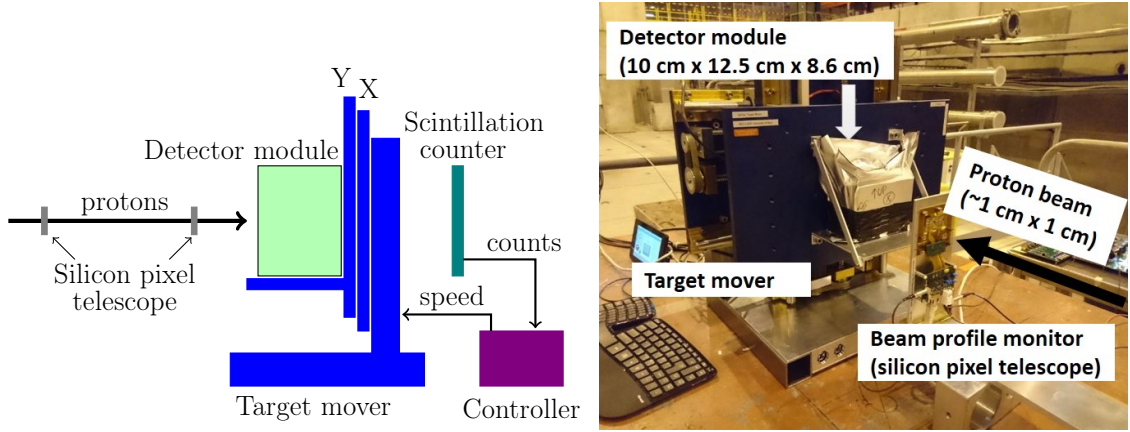
\begin{figure}[htpb]
\centering
\begin{tikzpicture}[scale=0.73,every node/.style={transform shape}]
\Large
\filldraw[fill=green!30, draw=black] (0, 1) rectangle (1.5,3);
\draw [ultra thick, ->] (-4.5,2) -- (-0.2,2); 
\draw [fill, blue] (-0.2, 0.75) rectangle (1.55,0.95);
\draw [fill, blue] (1.55, 0.25) rectangle ++(0.2,4);
\draw [fill, blue] (1.80, 0.0) rectangle ++(0.2,4);
\draw [fill, blue] (2.05, -0.5) rectangle ++(0.5,4);
\draw [fill, blue] (-0.5, -1) rectangle ++(3.5,0.6);
\draw [fill, gray] (-1.2, 1.75) rectangle ++(0.1,0.5); 
\draw [fill, gray] (-3.9, 1.75) rectangle ++(0.1,0.5); 
\draw [ <-] (-3.7,1.7) -- ++(0.4, -0.4); 
\draw [ <-] (-1.2,1.7) -- ++(-0.4, -0.4); 
\draw [fill, green!50!blue] (3.5, 1) rectangle ++(0.2,2);
\draw [fill, blue!50!red] (4, -1) rectangle ++(1.5,1); 
\draw [thick, ->] (3.7, 1.3) -- (5, 1.3) -- (5, 0.05); 
\draw [thick, ->] (4.2, 0) -- ++(0, 0.3) -- (2.6, 0.3); 
\node at (-2.4,2.3) {protons};
\node at (-0.3,3.3) {Detector module};
\node at (1.5,4.6) {Y};
\node at (2,4.4) {X};
\node (s) at (3.9,3.8) {Scintillation};
\node at (3.7,3.3) {counter};
\node (silicon) at (-2.4, 1.0) {Silicon pixel};
\node (silicon) at (-2.4, 0.5) {telescope};
\node at (1,-1.4) {Target mover};
\node at (4.75,-1.4) {Controller};
\node at (3.5, 0.6) {speed};
\node at (4.8, 1.6) {counts};
\end{tikzpicture}
\includegraphics[width=0.46\textwidth]{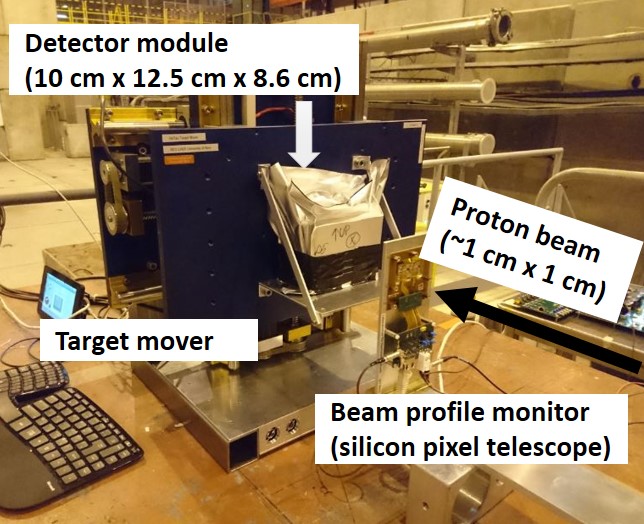}
\caption{Left: A schematic of DsTau setup. Right: Photo of the detector setup for the test beam campaign at the CERN SPS H4 beamline. The detector module was driven by a target mover so that it was uniformly exposed to the proton beam at a density of 10$^5$ protons/cm$^2$. 
}
\label{fig:test_beam}
\end{figure}

A schematic view of the detector setup is shown in Figure \ref{fig:test_beam}, and a photo of actual setup at the H4 beamline in November 2016 is also given. 

\begin{figure}[htbp]
    \centering
    \includegraphics[height=0.33\textwidth]{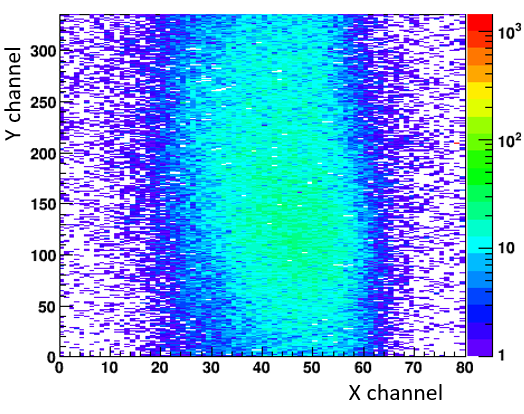}
    \includegraphics[height=0.33\textwidth]{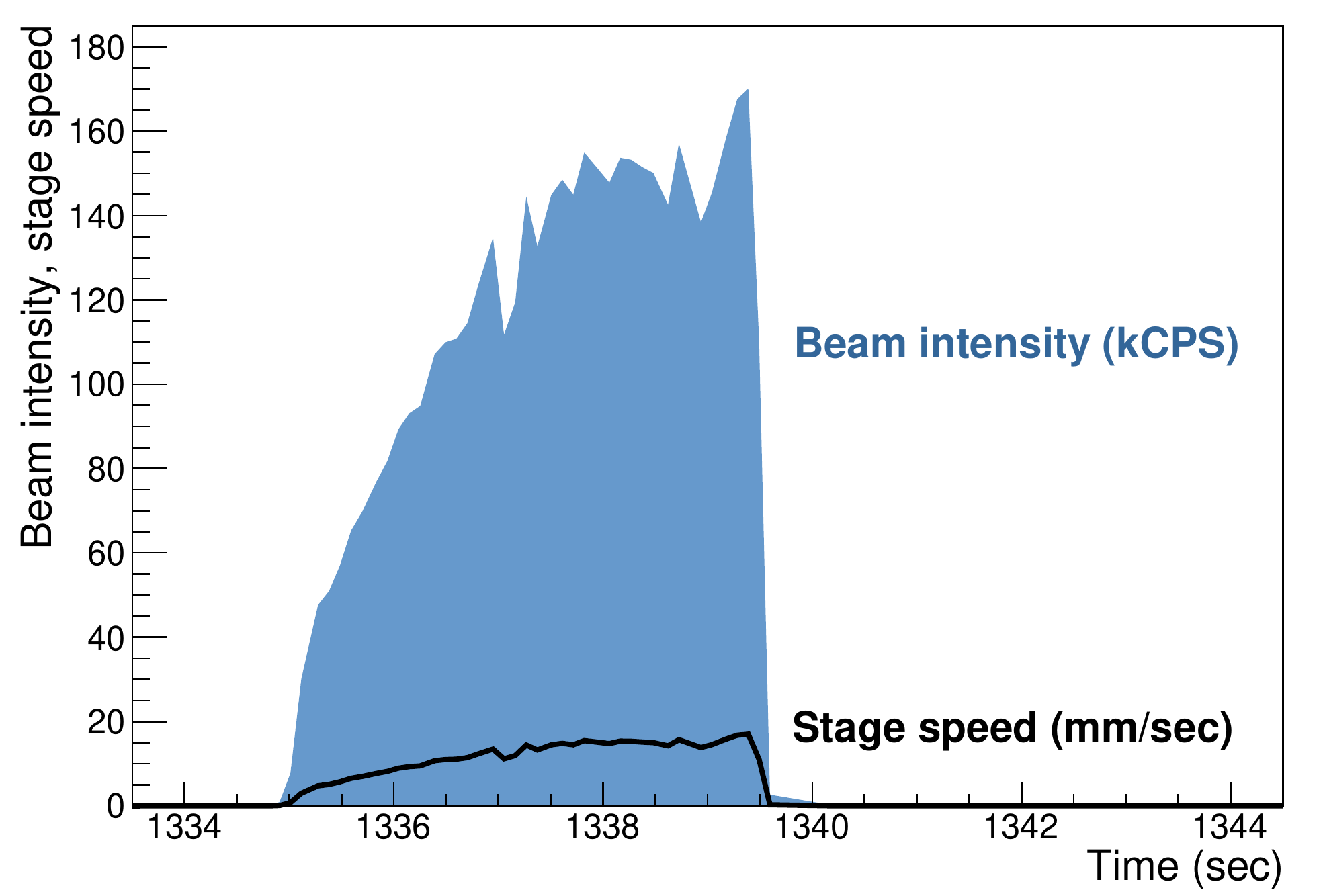}
    \caption{Left: A beam profile measured by the silicon pixel detector in 2016 beam test. The sensor area correspond to 2.0 cm$\times$1.9 cm. Right: The time structure of the proton beam measured by the scintillation counter, and the speed of target mover.}
    \label{fig:siprofile}
\end{figure}

The proton beam profile was measured by a silicon pixel telescope, which comprises two planes of the ATLAS IBL modules with FE-I4A front-end readout chips \cite{silicon_1,silicon_2}. The sensor dimension is 2.04 cm $\times$ 1.87 cm, it contains 80 $\times$ 336 pixels with a size of 250 $\mu$m $\times$ 50 $\mu$m.  To have uniform irradiation of the detector, the beam spot was enlarged to, for example, 1 cm$\times$ 2 cm  by tuning the beam optics, as shown in Figure \ref{fig:siprofile}.

\begin{figure}
    \centering
    \includegraphics[width=\textwidth]{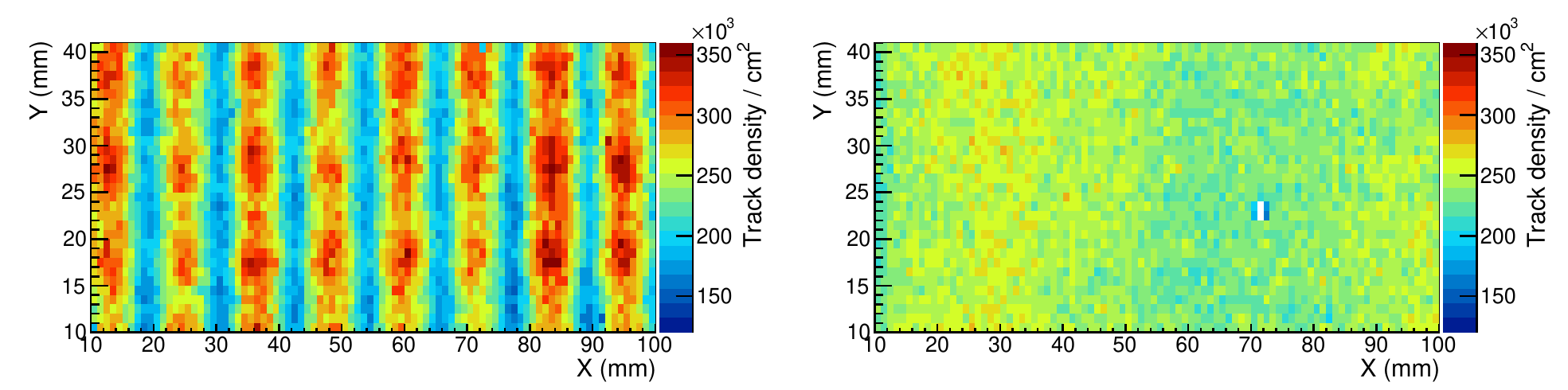}
    \caption{Track density distribution observed in the emulsion detector. Left: The target mover driven with a constant speed during beam on (2016). The periodical pattern corresponds to the scanning step in X. Right: The same plot with an intensity driven control (2017). }
    \label{fig:intensitydriven}
\end{figure}

Each emulsion detector module was mounted on a motorized X-Y stage (target mover) to change the position of the module with respect to the proton beam, so to make the detector surface uniformly irradiated at a density of \si{10^5\ tracks/cm^2}. 
The target mover consists of two axes driven by stepping motors and controlled by a Raspberry PI, which receives the SPS timing signal and the scintillation counter signal. During the beam spill, the target mover moves the detector in a horizontal direction perpendicular to the beam (X direction). When the end of the detector module in X is reached, the target mover comes back to the starting point in X with an increased vertical position (Y) by 1 cm.
A scintillation counter was implemented behind the target mover to monitor the rate of protons (See Figure \ref{fig:test_beam}). The discriminated signal was sent to a prescaler to reduce effective number of pulses, and then counted by Raspberry PI directly. The measured rate is fed into the target mover every 0.2 second and the speed in X direction was set proportional to the rate, as shown in Figure \ref{fig:siprofile} on the right. 
Figure \ref{fig:intensitydriven} on the right shows a comparison of the proton beam position profiles with a constant speed control during beam on and with the intensity-driven control. A better uniformity in proton irradiation was achieved with the latter one.

The exposure was done at room temperature. And then the modules were immediately cooled down in the fridge (5 \si{^\circ C}) to prevent the fading of recorded tracks. The development of films was done at the University of Bern. 

The emulsion detector is both a detector and the data storage media at the same time. 
To perform an efficient detection of the small kinks of $D_s\rightarrow\tau$ decays, the analysis is performed in two stages: (1) scan the full detector by a fast system with relatively coarse angular resolution and detect events those have two decays in a short distance, namely the decays of $\tau$ and partner charm ($D^\pm$ and $D^0$). (2) perform a high precision measurement around the $\tau$ decay candidates to find $D_s\rightarrow\tau$ small kinks.
The task (1) is performed by the fast scanning system, the HTS system. The angular resolutions of the track segments read out by the HTS is about 2.5 mrad for the angular range relevant for DsTau.

The second stage of the emulsions scanning (2) is performed at dedicated scanning stations which are very precise though not as fast as HTS. In these stations, a further improvement of the resolution of the track readout in the emulsion is achieved by replacing the conventional Z-axis stage (driven by a linear rail and stepping motor) by a piezo-based Z axis (PI PIFOC\textsuperscript\textregistered \ P-725.4CD
). This allows suppressing mechanical vibration in the horizontal plane, which limited the readout resolution in the conventional system. A reproducibility of single hit position measurement was achieved to be 8 nm. The angular measurement reproducibility is found to be 0.16 mrad (RMS), including measurement effects (such as skewing, hysteresis, encoding resolution of X/Y axes and optical aberrations). 

\subsection{Reconstruction and analysis}

The automatic scanning systems read out the track information accumulated in the emulsion film during the exposure, digitize it and transfer to the computers for the pattern recognition and track analysis like in case of any electronic detector. The output of the readout is the information on the track segments recorded in the top and bottom layers of the film (\textit{microtracks}) \footnote{Recall that a film is consists of two 70-\si{\micro m}-thick emulsion layers on both sides of a 210 \si{\micro m} plastic base.}. A segment made by linking the microtracks on two layers in a film is called a \textit{basetrack}, which is a basic unit of track information from each emulsion film for later processing. Each basetrack provides a 3D coordinates $\vec{X}=(x,y,z)$, 3D vector $\vec{V}=(\tan\theta_x,\tan\theta_y,1)$ and $dE/dx$ parameter.



The basic concept of track reconstruction is to link basetracks on different films by using their position and angular correspondences. The track density in DsTau ($10^5-10^6$/\si{cm^2} in small angular space) is relatively high. The conventional reconstruction tools for OPERA, which had a track density of $10^2-10^3$/\si{cm^2} in large angular space, are often not appropriate. 
A new tracking algorithm has been developed to overcome this problem of reconstruction in high track density in small angular space. More detail is given in Appendix \ref{appendix:algorithm}.

\begin{figure}[tb]
\centering
\includegraphics[height=5cm]{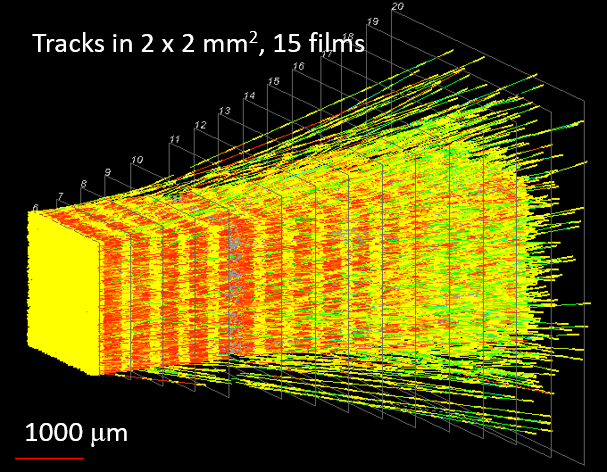}
\includegraphics[height=5cm]{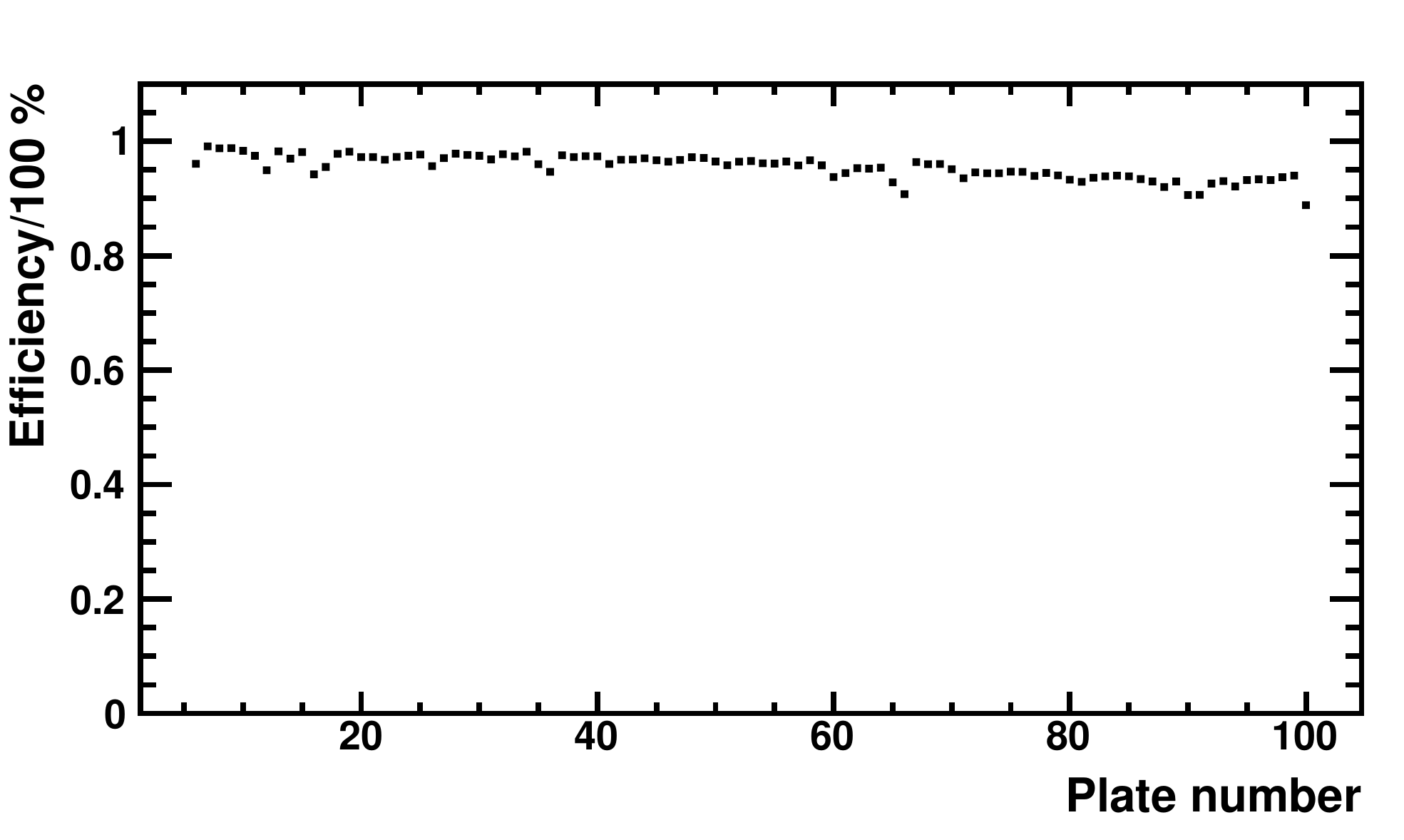}
\caption{Left: 3D display of reconstructed tracks. Right: Film-by-film basetrack efficiency in one of the decay modules in 2016 run.
}
\label{fig:tracks}
\end{figure}

An example of the reconstructed data from the detector is shown in the Figure \ref{fig:tracks} (left), which shows about ten thousand tracks in 2 mm$\times$ 2 mm ($\simeq 2.5\times 10^5$/\si{cm^2}).

The basetrack efficiency is measured by using reconstructed tracks and shown in Figure \ref{fig:tracks} (right). The average basetrack efficiency is higher than 95\%. Although it is a minor effect, there is a downward trend of the efficiency along the depth in the module. This is likely due to the increase of track density. The obtained basetrack efficiency is high enough to assure the track detection with high efficiency ($>$99\%).

\begin{figure}[htb]
    \centering
    \includegraphics[height=5cm]{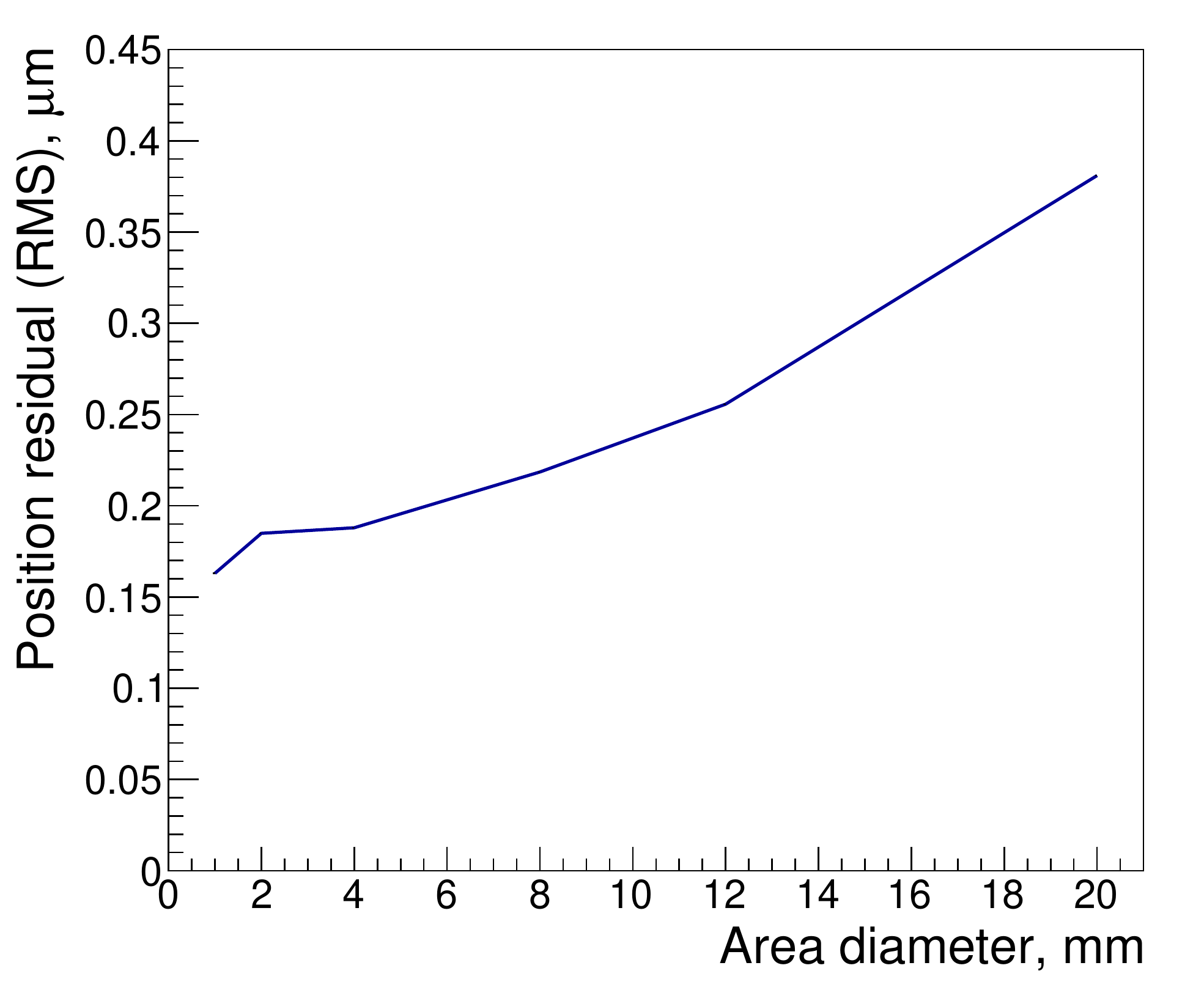}
    \includegraphics[height=5cm]{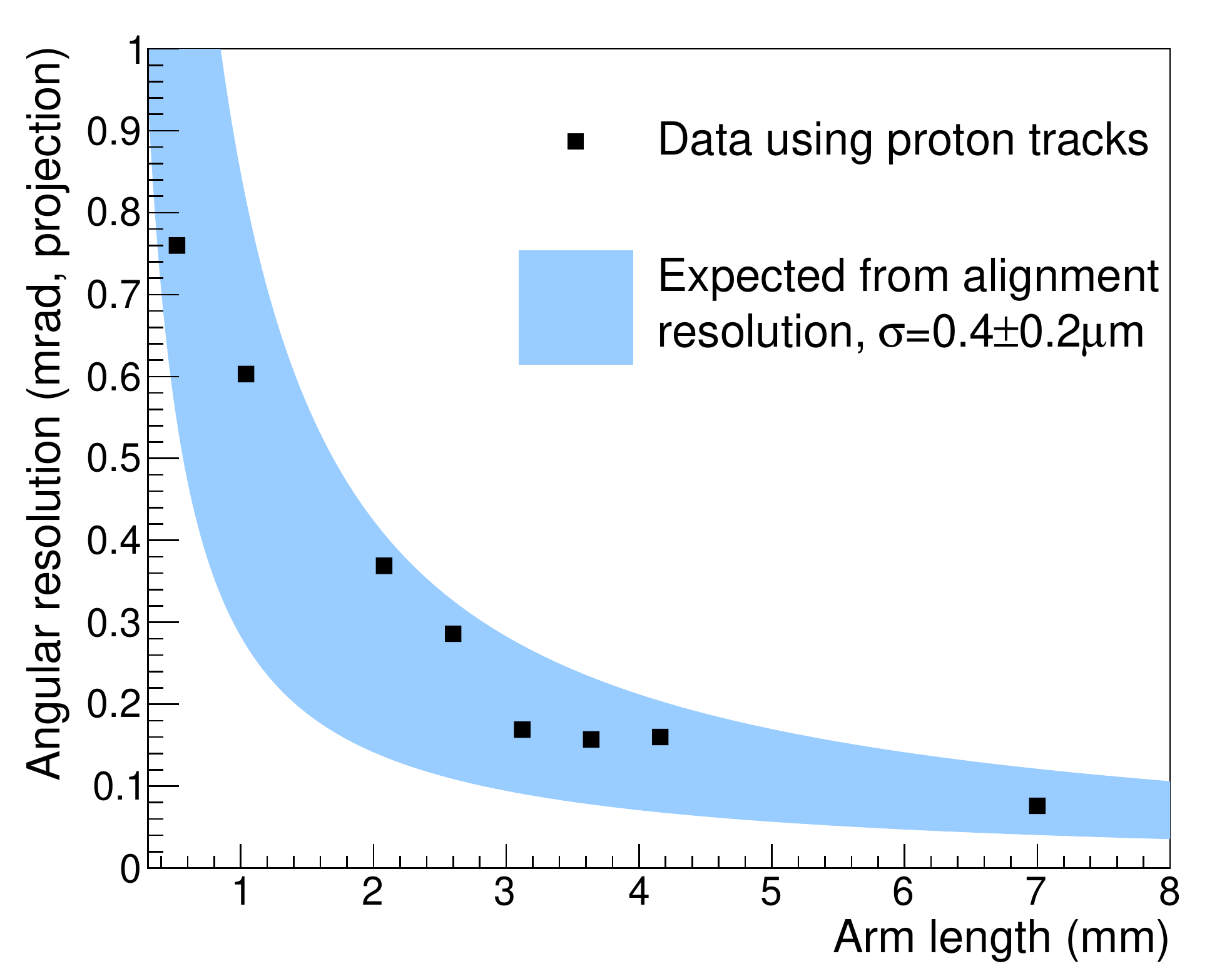}
    \caption{Left: Alignment resolution obtained in data as a function of the diameter of data processing unit. Position displacement of basetracks from the reconstructed track (a straight line) is evaluated. Right: Angular resolution of tracks as a function of reconstructed track length. The X-axis is the length used for the angular measurement.}
    \label{fig:alignment}
\end{figure}

The data processing is divided into sub-volumes, for example 2 cm $\times$ 2 cm $\times$ 15-30 emulsion films. 
An alignment between films (rotation, transverse position shifts and gap) is obtained by means of recorded tracks. In particular for the transverse position alignment, 400-GeV proton tracks those are supposed to be most straight are selectively used.
The obtained alignment precision with 15 films as a function of data processing unit diameter is shown in Figure \ref{fig:alignment} (left). The algorithm assures a sub-micron alignment in relatively large reconstruction unit. The alignment precision has a dependence on the processing area size, which is due to a non-linear distortion of the plastic base.

\begin{figure}[tb]
\centering
\includegraphics[width=0.32\textwidth]{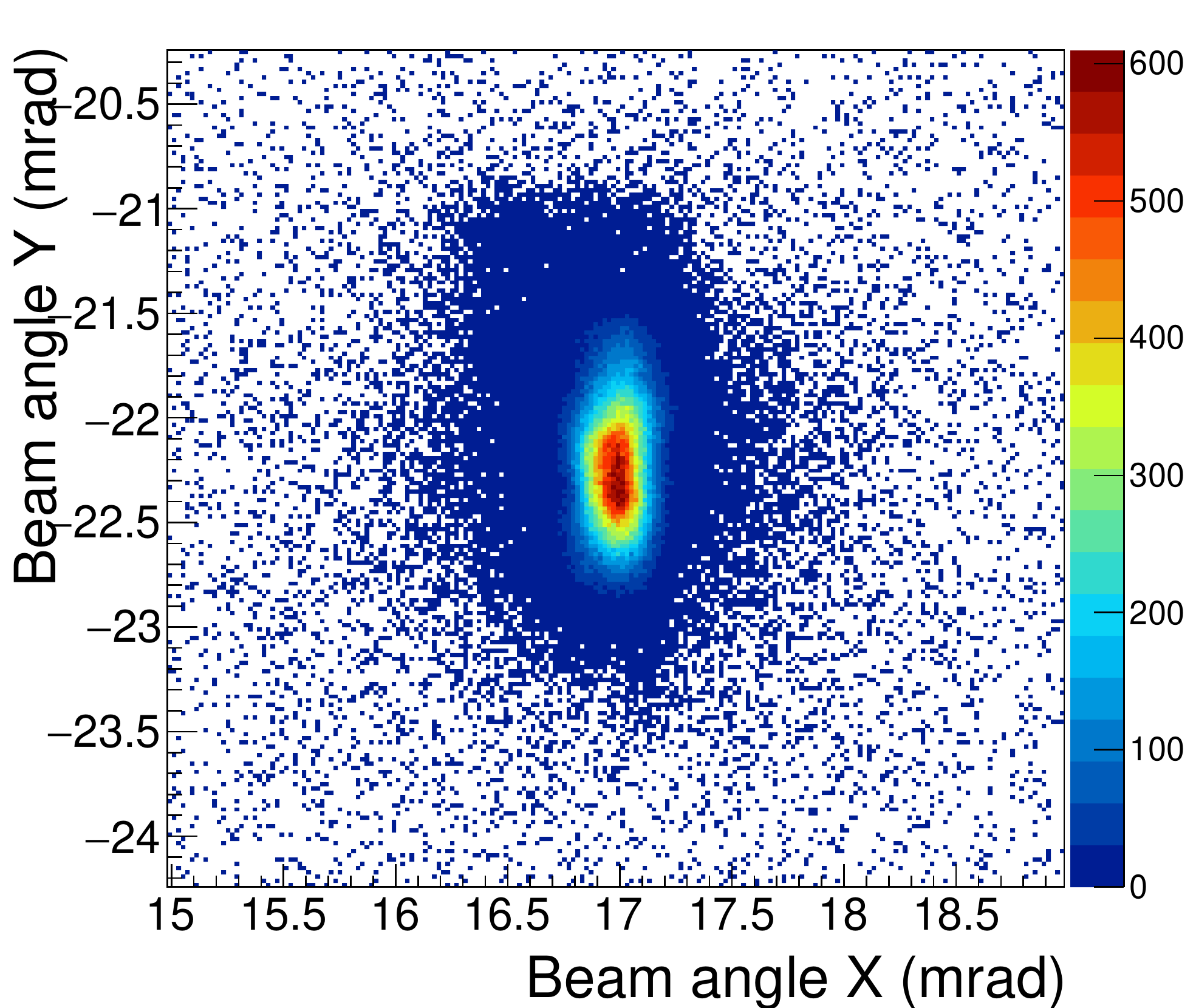}
\includegraphics[width=0.32\textwidth]{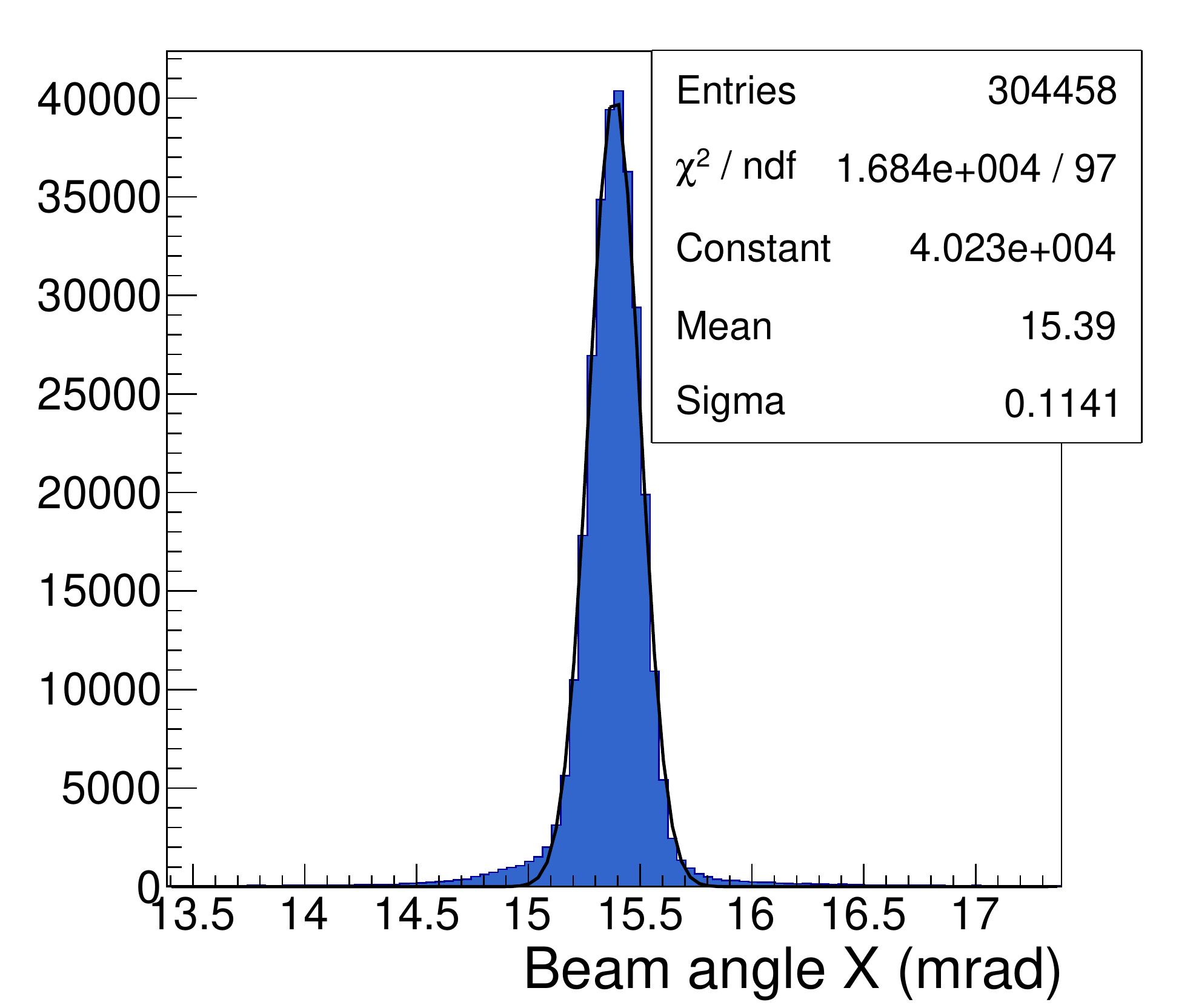}
\includegraphics[width=0.32\textwidth]{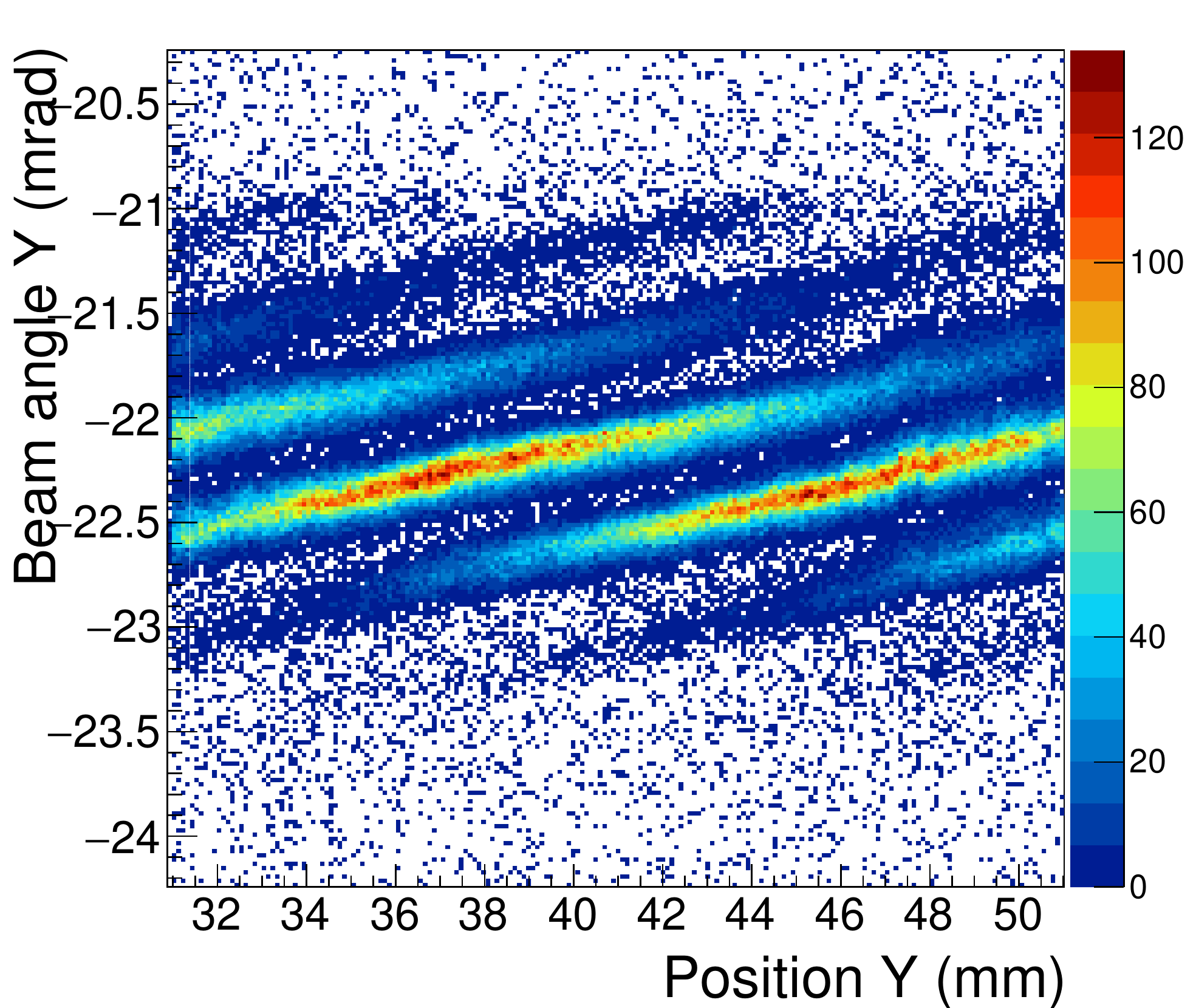}
\caption{Left: Angular distribution of proton tracks. The angular range is $\pm$2 mrad. Middle: X projection. The sigma of distribution is 110 \si{\micro rad}. Right: Beam angle Y as a function of Y coordinate. Each line corresponds to the exposure with a given Y coordinate. The positive correlation with position Y (slope of 0.44 mrad/cm) demonstrates the beam divergence.
}
\label{fig:beam}
\end{figure}

In order to test the quality of the alignment and reconstruction, the proton beam tracks were studied in detail. 
The beam angular distributions measured in 2 cm $\times$ 2 cm $\times$ 15 films (7 mm thick) are shown in Figure \ref{fig:beam} (left and middle). Approximated by Gaussian, the distributions have standard deviations of 110 \si{\micro rad} in XZ projection and 300 \si{\micro rad} in YZ. The width in angular distribution reflects the beam divergence. The beam position profile was wider in YZ projection (see Figure \ref{fig:siprofile}), therefore the beam divergence was also larger in YZ. In fact, Figure \ref{fig:beam} (right) is the Y position dependence of beam angle in YZ, showing a strong pattern. Each inclined line in this plot corresponds to exposure at a given Y position of the target mover. And the positive slope of the lines (440 \si{\micro rad/cm}) demonstrates that the beam is diverging. From the same plot, one can see  that the track angular resolution is better than 100 \si{\micro rad} in the large volume. When one of the lines is fitted, the standard deviation of track slopes from the fit was found to be 76 \si{\micro rad}. The angular resolutions with reconstructed tracks as a function of track length is shown in Figure \ref{fig:alignment} (right).


\begin{figure}[thb]
\centering
\includegraphics[width=0.56\textwidth]{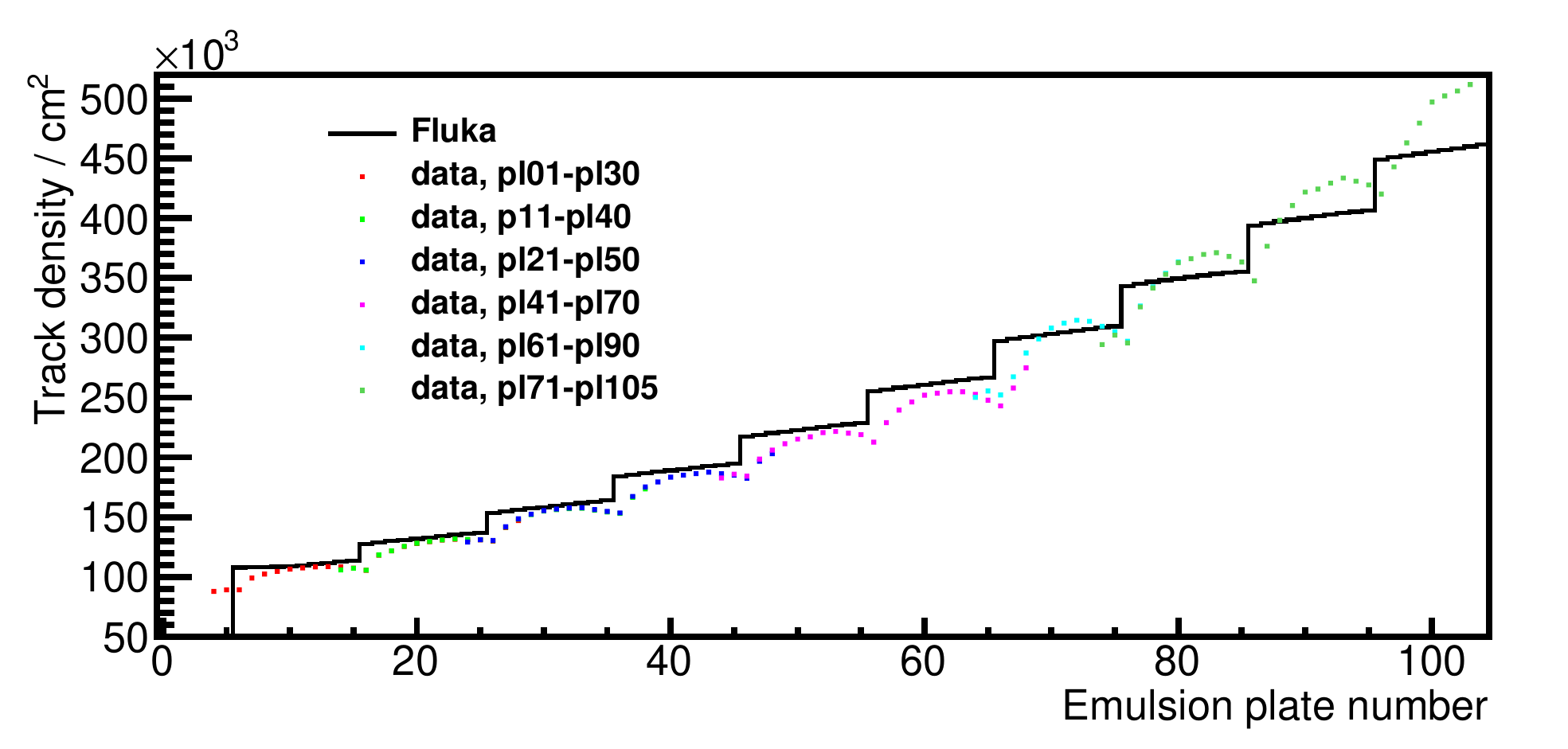}
\includegraphics[width=0.43\textwidth]{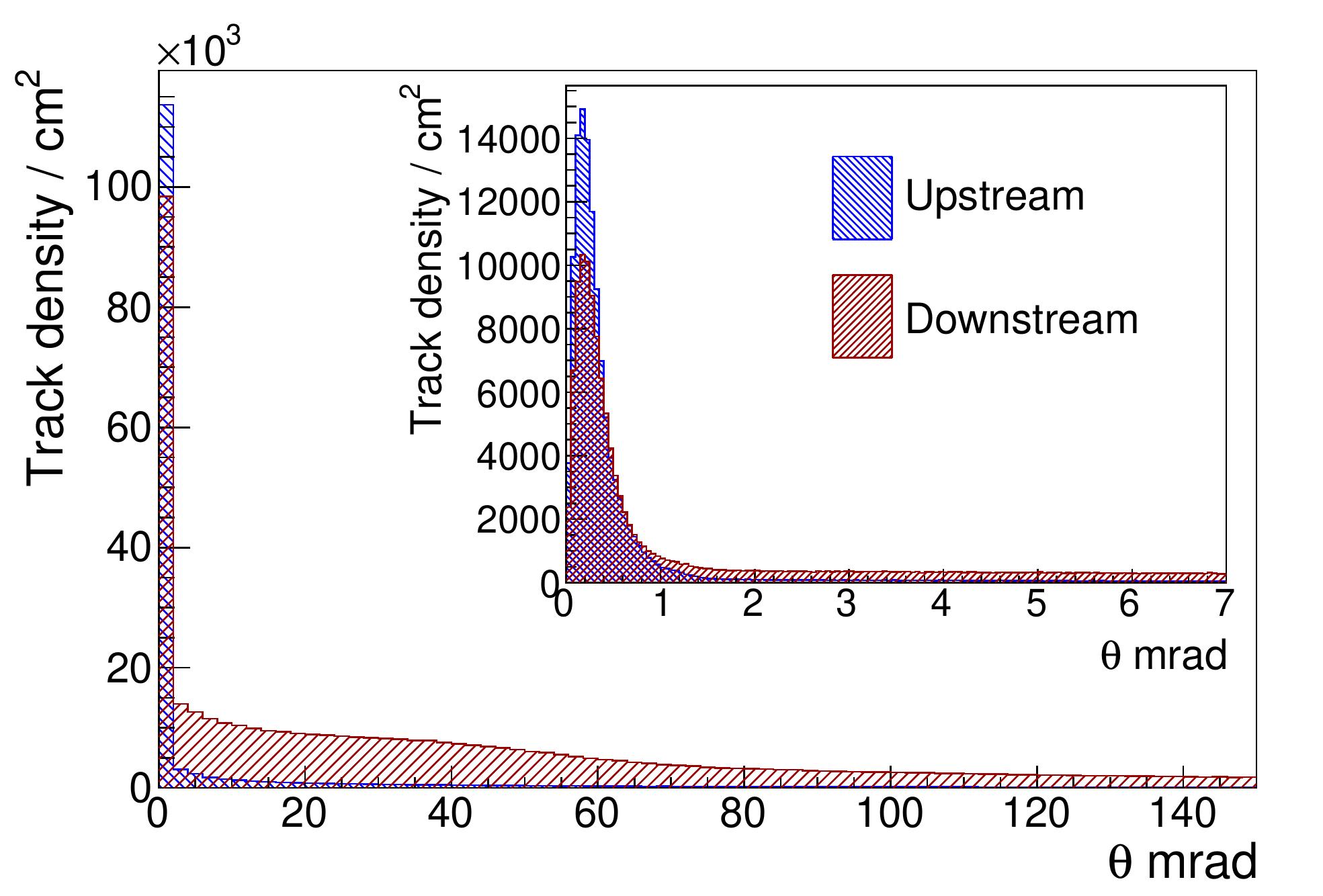}
\caption{Left: The evolution of track density as a function of depth in a module, compared with the FLUKA simulation (truth). Right: Comparison of track angular density (tracks/cm$^2$/bin) between an upstream unit and downstream. They correspond to plate number 6 and 86 in the left plot.
}
\label{fig:trackdensity}
\end{figure}

The track density in the detector module rises with the depth in the detector. The measured number of tracks is compared with the FLUKA simulation as shown in Figure \ref{fig:trackdensity} (left). 
The track density at the beginning of the decay module is about 1$\times 10^5$/\si{cm^2}, and increases up to $5\times 10^5$/\si{cm^2} in the last plates. This increase is due to the proton interaction daughters and their secondary interactions, as well as the electromagnetic showers.


Note that the total amount of material in the decay module corresponds to 0.1 $\lambda_{int}$ (0.05 by tungsten target, 0.05 by emulsion and plastic films), and 1.8 $X_0$ (1.4 $X_0$ by tungsten). 
Although the track density increases, it has a minor effect on the reconstruction since the measurements in the emulsion detector are not just points in space but micro-vectors with precise position and direction. The reconstruction quality (mis-connections, etc) would degrade if too many tracks will coincide both in their position and angular space. However, the secondary tracks in the detector have a large variety of angles, and the track density in angular space is not in fact increasing, as shown in Figure \ref{fig:trackdensity} (right). The plot shows the relative angle of tracks with respect to the proton beam angle. While the tail of relatively large angles increases, the track density in the region close to the beam direction ($\theta=0$) decreases because of proton interactions.

\begin{figure}[phtb]
\centering
\includegraphics[width=0.55\textwidth]{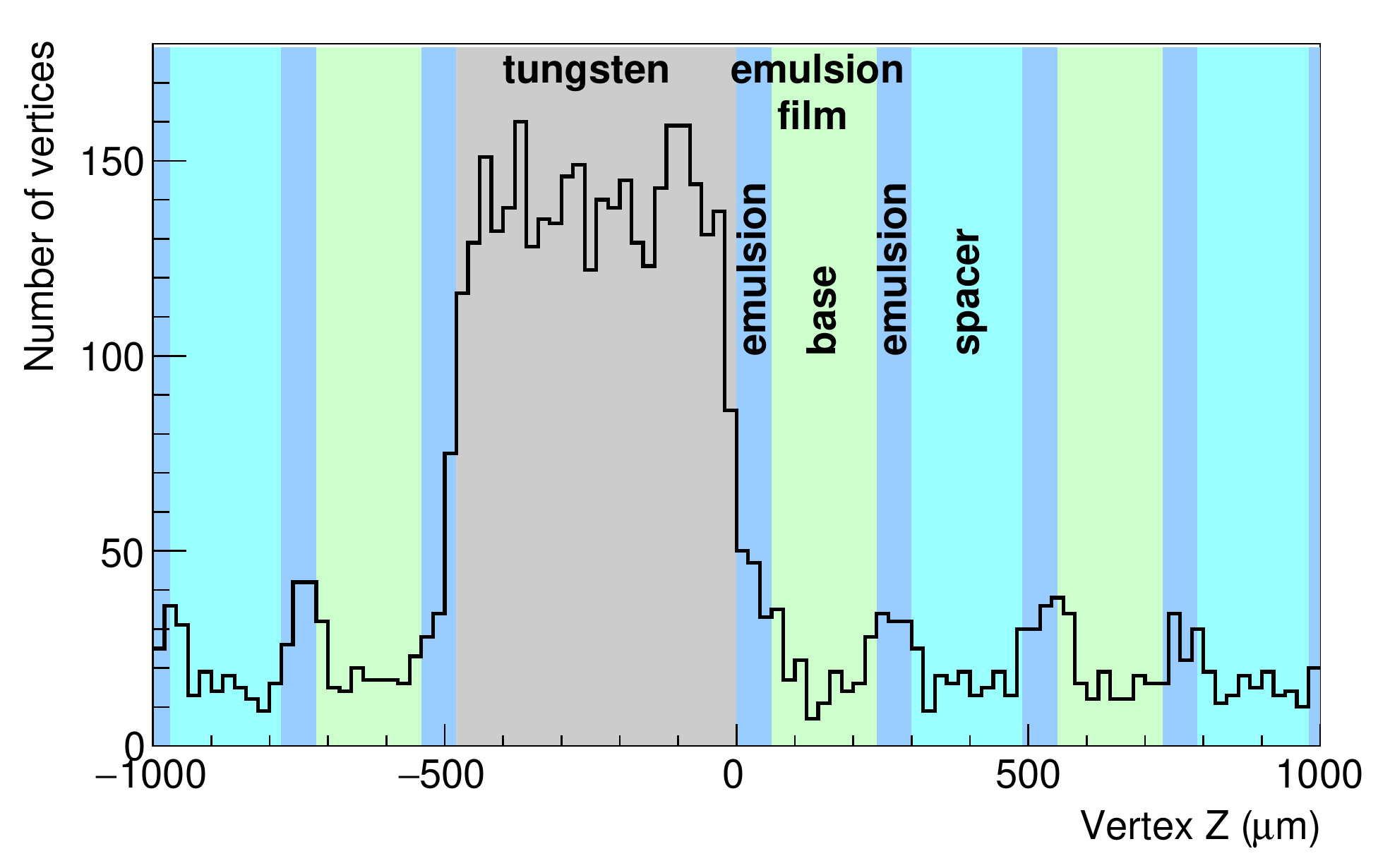}
\caption{Reconstructed vertex position distribution in Z. The correspondence with the detector structure is clearly visible.
}
\label{fig:vertices}
\end{figure}

\begin{figure}[phtb]
\centering
\includegraphics[width=0.55\textwidth]{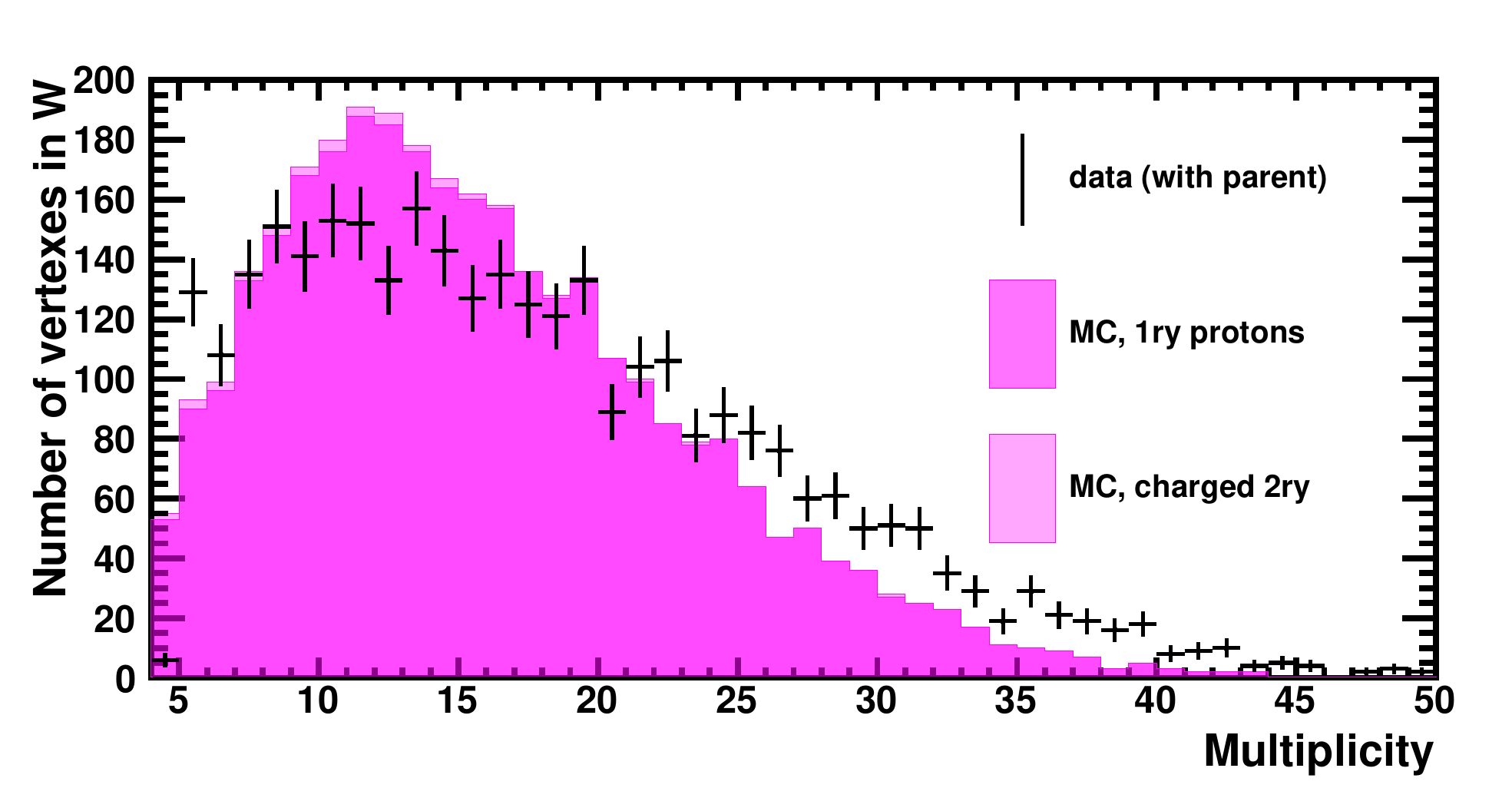}
\caption{Measured multiplicity of charged particles at the proton interaction vertices compared with the prediction from FLUKA simulations.}
\label{fig:multi_1ry}
\end{figure}

The reconstructed tracks are then used to find vertices. Using the tracks with the angle tan$\theta$ $<=$ 0.4 rad, conversing pattern with 4 or more tracks is considered as a vertex. Once a vertex is found, the parental proton track which has a minimum distance to the vertex within 3 \si{\micro m} is searched for. If the parent proton track is found, the vertex is considered as a primary proton interaction. 
Figure \ref{fig:vertices} shows a distribution of Z coordinate (along the beam) of the  vertexes reconstructed near by the tungsten target. An enhancement of vertices in the tungsten target is evident. One can even see the micro-structure corresponding to the emulsion layers (of higher density) and plastic bases/spacers. 
Figure \ref{fig:multi_1ry} shows the measured multiplicity of charged particles at proton interactions, compared with the prediction by FLUKA. Further study will be performed on the products of proton interactions.

\begin{figure}[htpb]
\centering
\includegraphics[width=0.9\textwidth]{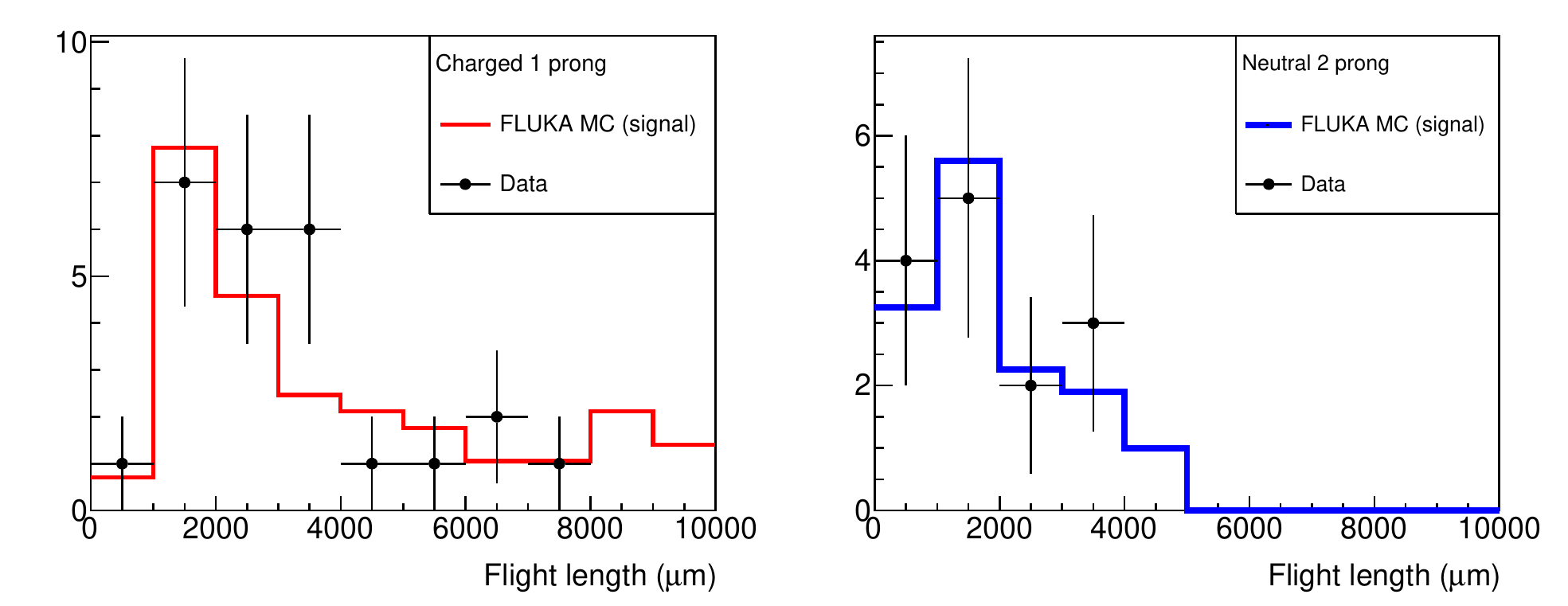}
\caption{Flight length distributions for charged 1-prong and neutral 2-prong decay candidates in the double-charm event samples. The FLUKA MC histograms are area-normalized to data.
}
\label{fig:FL_c1_and_n2_with_fluka}
\end{figure}

A systematic search for the decay topology of short-lived particles is applied to the found vertices. Events with decay topology are selected by applying 
the criteria equivalent to those described in Table \ref{tb:eff}.
The statistics of the found vertexes and events with the double decay topology observed in a sub-sample of the data are shown in Table \ref{tab:ds}. It is consistent with what is expected from the simulation for the equivalent data sample.   
The decay length distribution of the data and the FLUKA simulation is shown in Figure~\ref{fig:FL_c1_and_n2_with_fluka}, which demonstrates the exponential  behaviour of flight length expected for charmed particle decays.

\begin{table}[htb]
\centering
\begin{tabular}{|c|c|c|c|}
\hline
\ & Observed & \multicolumn{2}{c|}{Expected} \\
\hline
Vertices in tungsten & 29\,297  & \multicolumn{2}{c|}{$28\,390 \pm 910$ (syst)} \\
\hline
\ & \ & Signal & Background \\
\hline
Double decay topology & 20 & 13.5$\pm$3.2 & 2.4$\pm$0.3 \\
\hline
\end{tabular}
\caption{Statistics found in the sub-sample of data in which 5\,629\,670 protons are analyzed in one unit (1 tungsten plate). The systematic uncertainty on the expected number of vertices is due to the tungsten thickness fluctuation of 3.2\%.
}
\label{tab:ds}
\end{table}

\begin{figure}[htpb]
\centering
\includegraphics[width=0.7\textwidth]{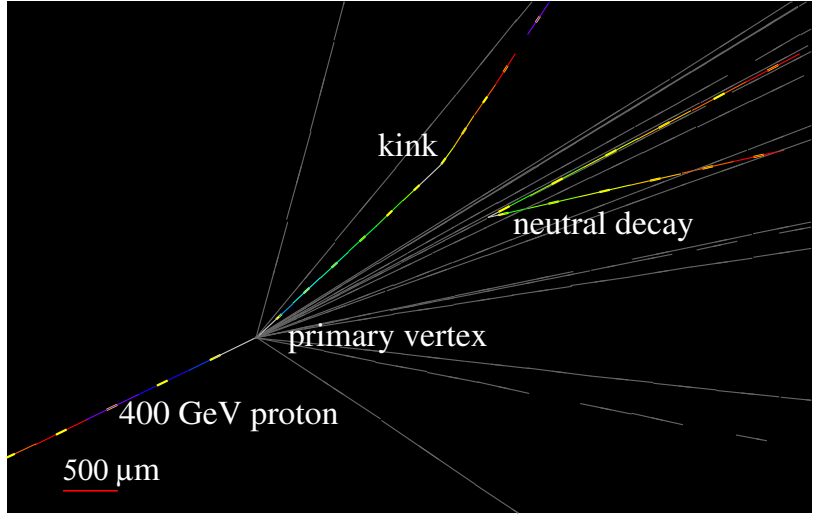}
\caption{A double charm candidate event with a neutral 2-prong (vee) and a charged 1-prong (kink) topology. (tilted view) See the text about the detail of event features.
}
\label{fig:doublecharm}
\end{figure}

An example of double charm event candidate found with the help of the analysis scheme is shown in Figure~\ref{fig:doublecharm}.
The proton interaction occurred in tungsten, 250 \si{\micro m} upstream of the following first emulsion film.
There are 18 charged particles at the proton interaction vertex ($v^\textnormal{1ry}$), conversing with a mean minimum distance to the vertex (impact parameter) of 1.6 \si{\micro m}. One of those has a kink at 3.32 mm from $v^\textnormal{1ry}$. The impact parameter of kink daughter to $v^\textnormal{1ry}$ is 174 \si{\micro m}. Additionally, two charged particles start 2.20 mm downstream of $v^\textnormal{1ry}$ with an opening angle of 0.132 radian. The plane made of these two particles has 15.2 mrad tilted with respect to the possible neutral parent vector, meaning that this neutral decay is not two body decay. It is unlikely that the neutral decay is due to $K_S^0$ because $K_S^0$ mostly decays into two body.
 


\section{Conclusion and outlook}
The DsTau experiment is going to study tau neutrino production following high energy proton interactions, which will provide necessary information for future $\nu_\tau$ experiments. 
The letter of intent and proposal of DsTau were submitted to CERN-SPSC \cite{dstau_loi, dstau_proposal} in 2016 and 2017, respectively. The SPSC recommended approving DsTau in April 2019.

The test beam in 2016-2017 and pilot run in August 2018 were performed, in which we accumulated over 20 million proton interactions in the detector. The emulsion  scanning and analysis of these sample are ongoing, which would allow to the re-evaluation of the $\nu_\tau$ cross-section by refining the $\nu_\tau$ flux. The first results presented here provide a convincing proof of the feasibility of the full scale study scheduled for the next physics run at CERN SPS in 2021.

\acknowledgments
We thank for the support of the beam physicists at CERN, in particular, A. Fabic, N. Charitonidis and B. Rae, and the SPS coordinator, H. Wilkens, for the successful data taking in the proton beam at the SPS north area. We would like to acknowledge the precious contributions of the technical staff from LHEP Bern. 
We acknowledge N. Naganawa and the colleagues from Nagoya University for sharing the expertise of emulsion detector. We warmly thank young students who helped the beam exposure campaign, A. Akmete, O. Durhan and  E. Elikkaya from Middle East Technical University, M. Morishita, Y. Nakamura and S. Tada, from Nagoya University.

This work was supported by JSPS KAKENHI Grant Numbers JP 18KK0085, JP 17H06926, and Konica Minolta Science and Technology Foundation Research Grant for Photographic Science. This work was partially supported by the Swiss National Science Foundation Ambizione grant PZ00P2\_154833, especially on the detector R\&D.

\appendix

\section{Review of the differential production cross section measurements for charm particles}
\label{appendix:charm}

The differential production cross section of $D_s$ in high-energy proton was rarely measured in previous experiments. The differential production cross sections of charmed mesons were conventionally approximated by a phenomenological formula, Eq. \ref{eq:differential_crosssection}. 
The longitudinal term is controlled by the parameter $n$ and the transverse term is done by $b$.
These parameters should be obtained experimentally. The HERA-B experiment \cite{Hera-B} used 920 GeV/c proton beam, and reported 18.5 $\pm$ 7.6 $\mu$b/nucleon using 11.4 $\pm$ 4.0 $D_s^+$ events. 
The Fermilab experiments E653 \cite{E653} and E743 (LEBC-MPS) \cite{E743} measured the $D$ meson production in 800 GeV/c proton interactions; the measured transverse dependence $b$ ($D^0$, $D^+$) = $0.84^{+0.10}_{-0.08}$ and $b$ ($D$) = 0.8 $\pm$ 0.2 were used in the DONUT analysis because the transverse dependence of charmed hadron production can be assumed to be the same for all charmed particles. 
The longitudinal dependence $n$ is known to be sensitive to the quark content as well as to the beam energy. E769 \cite{E769}, which used a 250 GeV proton beam, reported $n$ = 6.1 $\pm$ 0.7 for the inclusive charm meson production ($D^\pm$, $D^0$, $D_s^\pm$) \cite{E769}. All the other experiments using high-energy proton beams did not distinguish $D_s$ from all the other charmed particles ($D^\pm$, $D^0$, $\Lambda_c$) and only provided average values of $n$. Because the $D_s$ produced in proton-nucleus interaction does not contain valence quarks from the initial proton, the leading particle effect is reduced. Thus, the differential production cross section should be different from other charmed particles, which could contain quarks of incoming protons. E781 (SELEX) \cite{SELEX} used a 600-GeV $\Sigma^-$ beam (instead of a proton beam) and studied the difference between $D_s^+$ and $D_s^-$. The reported value for $D_s^+$, which is not affected by the leading particle effect, is $n$ = 7.4 $\pm$ 1.0 using about 130 $D_s^+$ events (within $x_F$ $>$ 0.15) in $\Sigma^-$ interactions (the value for $D_s^-$, which is affected by the leading particle effect, is $n$ = 4.1 $\pm$ 0.3). $D_s^+$ in SELEX may be similar to the DONUT situation; however, the incident particle is different, the measurement is limited to $x_F$ $>$ 0.15, and the uncertainty of $n$ is large owing to the limited data. These results are summarized in Table \ref{table_charm}. Results from the LHC experiments at $\sqrt{s}$ = 7, 8 or 13 TeV are not included here since the energies differ too much (400 GeV fixed target is at $\sqrt{s}$ = 27 GeV). LHCb started to measure charm productions in fixed target configuration ($D^0$ production was reported in \cite{lhcb_fixed}). LHCb might provide a good measurement also for $D_s$ in the future, however, the application of the LHCb measurements to the $\nu_\tau$ flux estimation is limited because the proton energy is very different (400 GeV w.r.t. 7-8 TeV); the uncertainty from the target atomic mass effect (17\% relative uncertainty in DONuT \cite{donut}) is large when extrapolating from their result with noble gas targets to tungsten one. 
In summary, no experimental result effectively constraining the $D_s$ differential cross section at the desired level, consequently the $\nu_\tau$ production, exists.

\begin{table}[t]
\centering
\scalebox{0.95}{
\scriptsize
\begin{tabular}{|p{2.4cm}|>{\centering}p{1.15cm}|>{\centering}p{1.35cm}|>{\centering}p{1.35cm}|>{\centering}p{1.35cm}|>{\centering}p{1.1cm}|p{3.9cm}|}
\hline
Experiment & Beam type / energy & $\sigma (D_s)$ (\si{\micro b/nucl})& $\sigma(D^\pm)$  (\si{\micro b/nucl})& $\sigma(D^0)$  (\si{\micro b/nucl})&$\sigma(\Lambda_c)$  (\si{\micro b/nucl}) & 
$x_F$ and $p_T$ dependence : \hspace{2cm}
$n$ and $b$ \si{(GeV/c)^{-2}}\\
\hline
HERA-B \cite{Hera-B} & p/920 & 18.4 $\pm$ 7.6 & 20.2 $\pm$ 3.7 & 48.7 $\pm$ 8.1 & - & $n(D^0, D^+) = 7.5 \pm 3.2$\\
\hline
Fermilab E653 \cite{E653} & p/800 & - & 38 $\pm$ 17 & 38 $\pm$ 13 & - & $n(D^0,D^+)=6.9^{+1.9}_{-1.8}$\\
& & & & & & $b(D^0,D^+)=0.84^{+0.10}_{-0.08}$\\ 
\hline
Fermilab E743 \cite{E743} & p/800 & - & 26 $\pm$ 8 & 22 $\pm$ 11& - & $n(D) = 8.6 \pm 2.0$ \\ 
(LEBC-MPS) & & & & & & $b(D) = 0.8 \pm 0.2$  \\ 
\hline
Fermilab E769 \cite{E769} & p/250 & 1.6 $\pm$ 0.8& 3 $\pm$ 1& 6 $\pm$ 2 & - & $n(D^\pm,D^0,D_s^\pm) = 6.1\pm 0.7 $\\ 
 & & & & & & $b(D^\pm,D^0,D_s^\pm) = 1.08\pm 0.09 $\\ 
\hline
Fermilab E769 \cite{E769} & $\pi^\pm$/250 & 2.1 $\pm$ 0.4& - & 9 $\pm$ 1 & - & $n(D^\pm,D^0,D_s^\pm) = 4.03\pm 0.18 $*\\ 
 & & & & & & $b(D^\pm,D^0,D_s^\pm) = 1.08\pm 0.05 $*\\ 
\hline
Fermilab E781  \cite{SELEX} & $\Sigma^-$/600  & - & - & - & - & $n(D_s^-) = 4.1 \pm 0.3$* \\ 
(SELEX) & & & & & & $n(D_s^+) = 7.4 \pm 1.0$*  \\ 
\hline
\end{tabular}
}
\caption{Charmed particle differential production cross section results obtained by the fixed-target experiments.  
*: incident particle is not proton.}
\label{table_charm}
\end{table}


\section{Tracking algorithm for reconstruction in high density environment}
\label{appendix:algorithm}

High resolution of the emulsion detector allows us to reconstruct proton interactions in an enormous pile-up of events or tracks of the order of $10^5$ tracks/\si{cm^2} in each film. This is also thank to the nature of data (basetracks, a track segments on a film), which has a vector information with  the position and angular resolutions of 0.4 \si{\micro m} and 2 mrad, respectively. 

The basic concept of track reconstruction is based on the correspondence of basetracks on different films in position and angular spaces. The widely-used algorithm uses a correspondence test of two consecutive basetracks. However, it can fail in the reconstruction in high density environment like in DsTau. 
Especially if two or more tracks with similar direction (within a few mrad) get close to each other (in a few micron), the algorithm may not resolve the correct paths.

\begin{figure}
      \centering
      \textbf{(a)} \includegraphics[width=0.8\textwidth]{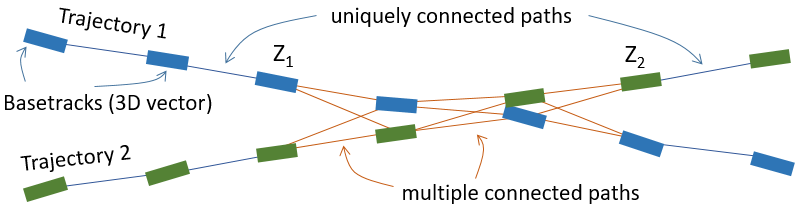}\\
      \textbf{(b)} \includegraphics[width=0.8\textwidth]{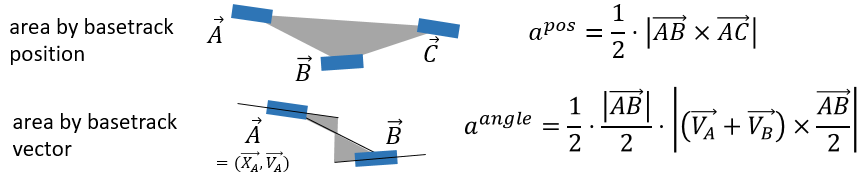}
      \caption{\textbf{(a)} Two tracks come closer and there are multiple possibilities in finding paths. \textbf{(b)} Areas made by basetracks, that are used to select the best possible paths.}
      \label{fig:algorithm}
  \end{figure}

New tracking algorithm was developed to reconstruct tracks in the environment with high track density and narrow angular spread. When it faces multiple path candidates, it keeps all possible paths. For the case in Figure \ref{fig:algorithm}-(a), $2^4=16$ possible paths are considered between  $Z_1$ to $Z_2$. For each path, it evaluates a test variable based on average of area made by basetracks involved in the possible paths,
\[
a^{average}=\left(\sum_i^{n-2}a^{pos}+\sum_i^{n-1} a^{angle}\right)/(n-0.5).
\]
Here, $a^{average}$ is an averaged area made by basetrack positions, $a^{pos}$, and one by basetrack angle, $a^{angle}$, as shown in Figure \ref{fig:algorithm}-(b). $n$ is the number of basetracks involved in the path and -0.5 is an empirical value to put a higher weight on longer paths. The path with the smallest $a^{average}$ is chosen to be the best one. This procedure is repeated by removing the basetracks involved in the chosen paths.

\end{document}